\documentclass[conference
]{IEEEtran}
\IEEEoverridecommandlockouts    
\usepackage{verbatim} 
\usepackage{cite}
\usepackage{nicematrix}
\usepackage{longtable}
\usepackage{booktabs}
\usepackage{bm}
\usepackage{amsmath,amssymb,amsfonts,flushend}
\makeatletter
\let\MYcaption\@makecaption
\makeatother
\usepackage[font=footnotesize]{subcaption}
\makeatletter
\let\@makecaption\MYcaption
\makeatother
\usepackage{algorithmic}

\usepackage{graphicx}
\usepackage{tikz}
\usepackage{circuitikz}
\usepackage{siunitx}
\usetikzlibrary{shapes.geometric, arrows,chains, positioning}
\usetikzlibrary{arrows.meta} 
\usetikzlibrary { decorations.pathmorphing, decorations.pathreplacing, decorations.shapes, } 
\usepackage{pbox}
\usepackage{relsize}
\tikzset{
    pin/.style = {font = \relsize{-2}} 
}
\ctikzset{
    bipoles/length = 4em, 
    font = \relsize{-1}, 
}
\usepackage{textcomp}
\usepackage{xcolor}
\def\BibTeX{{\rm B\kern-.05em{\sc i\kern-.025em b}\kern-.08em
    T\kern-.1667em\lower.7ex\hbox{E}\kern-.125emX}}
\usepackage{mathtools}
\usepackage{schemabloc}
\usetikzlibrary{circuits}
\usetikzlibrary{arrows}

\usepackage{pgfplotstable}
\usepackage{multirow}
\usepackage{mathtools}
\usepackage{commath}

\newcommand{\RNum}[1]{\uppercase\expandafter{\romannumeral #1\relax}}
\graphicspath{ {./confimages/} }
 \usepackage{pgfplots}
  \pgfplotsset{compat=newest}
  \usetikzlibrary{plotmarks}
  \usetikzlibrary{arrows.meta}
  \usepgfplotslibrary{patchplots}
  \usepackage{grffile}
  \usepackage{amsmath}
  
\usepackage{amsthm}

\usepackage{fancyhdr,graphicx,amsmath,amssymb}
\usepackage[ruled,vlined]{algorithm2e}

\usepackage{times}
\usepackage{hhline}

\usepackage{hhline}
\usepackage{booktabs} 
\usepackage{circuitikz}
\usepackage{siunitx}

\usepackage[hyphens]{url}
\usepackage{balance}
\usepackage[commandnameprefix=always]{changes}
\definechangesauthor[color=blue,name=Vassilis]{VP}


\begin{document}

%
\allowdisplaybreaks 

\title{Evaluating Beam Sweeping for AoA Estimation with an RIS Prototype: Indoor/Outdoor Field Trials\vspace{-0.3cm}}

\author{\IEEEauthorblockN{ 
Dimitris Vordonis\IEEEauthorrefmark{2}, Dimitris Kompostiotis\IEEEauthorrefmark{2}, Vassilis Paliouras\IEEEauthorrefmark{2}, George C. Alexandropoulos\IEEEauthorrefmark{5}, and Florin Grec\IEEEauthorrefmark{4}}
\IEEEauthorblockA{\IEEEauthorrefmark{2}Electrical and Computer Engineering Department, University of Patras, Greece,\\
 \IEEEauthorrefmark{5}Department of Informatics and Telecommunications, National and Kapodistrian University of Athens, Greece,\\
\IEEEauthorrefmark{4}European Space Agency (ESA), Noordwijk, Netherlands\\
e-mails:\{d.vordonis, d.kompostiotis\}@ac.upatras.gr, paliuras@upatras.gr, alexandg@di.uoa.gr, Florin-Catalin.Grec@esa.int}\vspace{-1.07cm}}

\markboth{NOTES}%
{Shell \MakeLowercase{\textit{et al.}}: Bare Demo of IEEEtran.cls for IEEE Journals}

\maketitle


\begin{abstract}
Reconfigurable Intelligent Surfaces (RISs) have emerged as a promising technology to enhance wireless communication systems by enabling dynamic control over the propagation environment. However, practical experiments are crucial towards the validation of the theoretical potential of RISs while establishing their real-world applicability, especially since most studies rely on simplified models and lack comprehensive field trials. In this paper, we present an efficient method for configuring a 1-bit RIS prototype at sub-6 GHz, resulting in a codebook oriented for beam sweeping; an essential protocol for initial access and Angle of Arrival (AoA) estimation. The measured radiation patterns of the RIS validate the theoretical model, demonstrating consistency between the experimental results and the predicted beamforming behavior. Furthermore, we experimentally prove that RIS can alter channel properties and by harnessing the diversity it provides, we evaluate beam sweeping as an AoA estimation technique. Finally, we investigate the frequency selectivity of the RIS and propose an approach to address indoor challenges by leveraging the geometry of environment.
\end{abstract}


\section{Introduction}
Reconfigurable Intelligent Surfaces (RISs) are emerging as a transformative technology for 6G~\cite{huang2019reconfigurable,bjornson2022reconfigurable,pan2022overview}, offering the ability to dynamically control wireless propagation environments with minimal energy consumption. In the context of wireless communication systems, RISs hold great promise for positioning applications~\cite{bjornson2022reconfigurable,pan2022overview,wymeersch2020radio,wymeersch2022radio,keykhosravi2023leveraging,alexandropoulos2023ris}, enabling precise user localization even in challenging scenarios such as urban areas with blockages or dense multipath. By introducing additional degrees of freedom through controllable reflections, RISs can improve the Signal-to-Noise Ratio (SNR), which is critical for positioning algorithms, while leveraging their large aperture size and acting as extra reference points to enhance localization accuracy. Furthermore, RISs can exploit and control multipath propagation, turning it into a valuable resource by providing diverse and rich signal measurements to aid positioning systems.

Building on the ability of RIS to improve wireless positioning, efficient beamforming strategies are essential to fully exploit its capabilities. One such practical and widely adopted protocol in wireless communications is beam sweeping~\cite{che2023efficient,10048686,ouyang2023computer}, which plays a critical role in both initial access and Angle of Arrival (AoA) estimation.
In RIS-aided systems, an RIS node enables beam sweeping by steering the incoming signal and reflecting it toward the desired angles, systematically scanning the served area with narrow beams. To reduce resource usage and pilot signaling, variations such as hierarchical codebooks~\cite{albanese2021papir,RIS_hierarchical}, multi-beam patterns~\cite{wang2023hierarchical}, or broad beams~\cite{ramezani2023dual,ramezani2024broad,10256051,lin2024design} can be used, offering efficient coverage and supporting user mobility.


 
Despite its great potential, the RIS technology necessitates rigorous experimental validation to substantiate its theoretical benefits. The imperfections of RIS prototypes significantly complicate the design of effective codebook-based beamforming strategies. Issues such as strong side lobes in radiation patterns of 1-bit RIS designs~\cite{10498085,sayanskiy20222d}, phase-dependent amplitude response~\cite{10095408,kompostiotis2023secrecy}, frequency selectivity~\cite{rains2023fully}, dependence on the angle of incoming signals~\cite{9551980,weinberger2024validating}, and the practical difficulty of achieving precise binary phase shifts~\cite{9551980,rains2023fully}, meeting the $180^\circ$ difference, are critical limitations. Furthermore, mutual coupling between adjacent RIS elements~\cite{zheng2024mutual,9837634}, limited scanning range, and the impracticality of achieving narrow beams at certain angles with 1-bit resolution~\cite{dai2020reconfigurable}, further degrade performance. These hardware-related imperfections, combined with environmental challenges such as multipath propagation~\cite{kompostiotis2024evaluation,nikonowicz2024indoor}, ground-bounce effects~\cite{trichopoulos2022design}, and interference, make the task of experimental validation even more difficult. Overcoming these challenges is crucial to ensure the practical viability of RIS-aided systems in real-world deployments.

Recent measurement-based studies have focused mainly on evaluating SNR improvements and understanding the radiation patterns of RIS~\cite{9551980,amri2021reconfigurable,dai2020reconfigurable,trichopoulos2022design}, particularly highlighting the presence of strong side lobes in low-resolution phase configurations. Some works aim to derive theoretical path loss models~\cite{tang2022path,9713744}, which have been validated through controlled experiments in anechoic chambers. However, extending these models to real outdoor scenarios~\cite{lan2023measurement} has revealed discrepancies caused by hardware imperfections and challenges inherent to real-world deployments. In this paper, we validate a subcase~\cite{ramezani2023dual,ramezani2024broad,10256051} of the model~\cite{9713744} that eliminates the need for approximations of transmitter (Tx) or receiver (Rx) antenna radiation patterns, simplifying real-world deployment considerations. Beyond system-level performance, RISs have gained significant attention for positioning applications. Ouyang et al.~\cite{ouyang2023computer} combine beam sweeping with visual information from a camera attached to the RIS to improve positioning adaptively in dynamic environments. Rahal et al.~\cite{rahal2023ris} integrate beam sweeping in combination with the Space-Alternating Generalized Expectation-maximization (SAGE) algorithm to mitigate side lobe challenges and multipath interference, refining position estimates in complex indoor settings. In contrast, our approach leverages beam sweeping, not only with offline codebooks, but also introduces a real-time configuration methodology that exploits the indoor environment's geometry and leverages non line-of-sight (NLoS) components. In addition, we propose an alternative beam sweeping strategy that minimizes the received signal power; an extra solution that complements the traditional objective of maximizing signal strength. 
Spatial Time Coding Modulation (STCM) has emerged as another method for AoA estimation~\cite{gholami2024wireless,dos2024assessing}, where RIS elements are time-modulated to generate harmonics that encode spatial information. However, STCM faces significant practical challenges~\cite{gholami2024wireless,dos2024assessing}, including strict timing constraints and phase synchronization issues, which are exacerbated by RIS configuration mechanisms, such as script-based control methods that introduce timing discrepancies, making it difficult to achieve the precise modulation intervals required for clean harmonics and reducing the accuracy of AoA estimation. In contrast, beam sweeping offers a simpler and more practical solution.




The remainder of the paper is organized as follows: Section~\ref{s:theoretical_model} introduces the theoretical model for the radiation pattern of RIS, which is validated through real outdoor measurements. Section~\ref{s:phase_optimization} presents one phase optimization method to configure the RIS prototype during the experiments. Section~\ref{s:measurement_setup} describes the measurement setup, including both outdoor and indoor environments, as well as the key system components. Then in Section~\ref{s:experimental_results}, we make several experiments
 to verify the system performance. This section further introduces the proposed beam sweeping methodology and evaluates its performance in both outdoor and indoor environments. Finally, Section~\ref{s:conclusion} concludes the paper.

\section{Radiation Pattern of RIS: Theoretical model}\label{s:theoretical_model}
\vspace{0.05cm}
We consider a single-antenna Tx, an RIS with $N {=} N_x {\times} N_y$ elements, where $N_x$ and $N_y$ denote the number of rows and columns, respectively, and a single-antenna Rx. The received signal, over a single subcarrier, can be modeled as in~\cite{rabault2024tacit}:
\begin{equation}
    y =  (\bm{h}_{r}^\mathsf{H}\bm{\Theta}\bm{h}_{g} + h_{d})s + z, 
    \label{e:rx_signal_A}
\end{equation}
where \( \bm{h}_{g},\bm{h}_{r} {\in} \mathbb{C}^{N \times 1} \) 
are the channels from the Tx to the RIS and from the RIS to the Rx, respectively; \( h_{d} {\in} \mathbb{C} \) is the direct link channel between the Tx and Rx; and \( s \) is the transmitted symbol with zero-mean and unit variance \( \mathbb{E}(|s|^2) {=} 1 \). The RIS is configured with a diagonal phase-shift matrix
$ \bm{\Theta} = \operatorname{diag}(\bm{\omega_\theta}^\mathsf{H})$, where $\bm{\omega_\theta} = [e^{\text{j}\theta_{1}}, \dots, e^{\text{j}\theta_{N}}]^\mathsf{T}$ and \( \theta_n \) represents the phase shift applied by the \(n\)-th RIS element. As \eqref{e:rx_signal_A} shows, \( \bm{\Theta} \) directly influences the received signal, allowing the RIS to manipulate the end-to-end channel. 

Regarding the theoretical model of the RIS radiation pattern~\cite{ramezani2023dual,ramezani2024broad,10256051}, it represents a specific subcase of the general model introduced in~\cite{9713744}, eliminating the need for approximations of the Tx or Rx antenna radiation patterns. This simplification is particularly valuable for facilitating practical deployment in real-world scenarios. The overall radiation pattern \( G(\phi,\theta) \) is expressed as~\cite{ramezani2024broad,10256051},
\begin{equation}
    G(\phi,\theta) = A(\phi,\theta) G_{R0}(\tilde{\phi}, \tilde{\theta}) G_{R0}(\phi,\theta),
    \label{eq:radiation_pattern}
\end{equation}
where \( (\tilde{\phi}, \tilde{\theta}) \) denote the AoA at the RIS, \( (\phi, \theta) \) are the angles of departure (AoD) from the RIS, and \( G_{R0} \) represents the radiation pattern of a single RIS element, modeled using the 3GPP antenna gain model~\cite{ramezani2024broad,10256051}. The power-domain array factor is given by,
\begin{equation}
    A(\phi, \theta) = \big| \bm{\omega}_{\bm{\theta}}^T \big(\mathbf{a}_{\text{RIS}}(\tilde{\phi},\tilde{\theta}) \odot \mathbf{a}^{*}_{\text{RIS}}(\phi, \theta)\big) \big|^2,
    \label{eq:power_factor}
\end{equation}
where \( \odot \) denotes the Hadamard product, and \( \mathbf{a}_{\text{RIS}}(\phi, \theta) \) is the array response vector for the RIS. For an RIS element located at \( \mathbf{u}_i \), \( \mathbf{a}_{\text{RIS}}(\phi, \theta) \) is defined as~\cite{ramezani2024broad,10256051}:
\begin{equation}
    \mathbf{a}_{\text{RIS}}(\phi, \theta) = \Big[e^{-\jmath\mathbf{k}(\phi, \theta)^\mathsf{T}\mathbf{u}_1}, \dots, e^{-\jmath\mathbf{k}(\phi, \theta)^\mathsf{T}\mathbf{u}_N}\Big]^\mathsf{T},
    \label{e:array_response_vec}
\end{equation}
where \( \mathbf{u}_i \) is the position of the \( i\)-th RIS element, and \( \mathbf{k}(\phi, \theta) \) represents the wave vector of the transmitted signal:
\begin{equation}
    \mathbf{k}(\phi, \theta) = \frac{2\pi}{\lambda} \Big[\cos(\theta)\cos(\phi), \cos(\theta)\sin(\phi), \sin(\theta)\Big]^\mathsf{T},
    \label{e:wave_vector}
\end{equation}
with \( \lambda \) being the wavelength.

\section{Practical Phase Optimization Method}\label{s:phase_optimization} 
The $\text{A}(\varphi,\theta)$ is maximized for
\begin{equation}
    \bm{\omega}_{\bm{\theta}}^{T} = (\mathbf{a}_{\text{RIS}}(\varphi_{\text{AoA}},\theta_{\text{AoA}}) \odot \mathbf{a}^{*}_{\text{RIS}}(\varphi,\theta))^{H}.
    \label{opt_config}
\end{equation}
in 
the continuous case, where each RIS element achieves infinite angular resolution. However, practical RIS implementations often restrict each element to discrete phase-shift values, such as \( \theta_i {\in} \{-\frac{\pi}{2}, \frac{\pi}{2}\}, \forall i \). Quantizing the optimal continuous solution (1-bit resolution) does not guarantee that the RIS will steer the beam toward the desired AoD and results in strong sidelobes~\cite{10498085,sayanskiy20222d}, which can significantly degrade performance. Therefore, this section presents a practical RIS phase optimization method based on column-row scanning. Grouping elements, such as by columns or rows, is a well-established approach to reduce complexity, leveraging the fact that adjacent elements typically exhibit similar channel coefficients~\cite{bjornson2021optimizing}. A similar idea to column/row changes is discussed in~\cite{9551980}. Iterative algorithms that optimize each element individually often require more iterations and, in long-range scenarios without low-noise amplifiers (LNAs), small differences in received signal power may be challenging to detect~\cite{weinberger2024validating,tewes2023comprehensive}. The applied method starts with a homogeneous surface, initializing all elements of the RIS to the same configuration. The optimization proceeds horizontally by modifying the phase-shift states of individual columns, measuring the received power at the Rx. If an improvement is observed, the change is retained; otherwise, the configuration is reverted. Next, a similar procedure is applied vertically across rows. This process runs for one complete iteration, providing a low-complexity yet effective configuration strategy. The complete procedure is detailed in Algorithm~\ref{algo:ris_config}.
\begin{algorithm}[b]
\footnotesize
\parbox[t]{0.7\textwidth}{\caption{\strut\small\textbf{RIS Configuration via Column-Row Scanning\vskip-5ex}} \label{algo:ris_config}}
\tcp{\( \bm{\omega_\theta} \) is mapped to the dual-polarized RIS's configuration matrix \( \bm{\Phi} \) of size \( N_x \times 2N_y \).}
\KwIn{Initial RIS configuration matrix \( \bm{\Phi} \) with all elements set to 0.}
\KwOut{Optimized RIS configuration \( \bm{\Phi} \).}
\nl Measure the initial received power \( P_{\text{max}} \) with \( \bm{\Phi} \) set to all 0\;
\nl 
    \tcc{\textbf{Step 1: Column-Wise Scanning}}
    \nl \For{\texttt{\( col = 1:2:2N_y \)}}{
        \nl Invert the states of columns \( col, col+1 \) in \( \bm{\Phi} \)\;
        \nl Measure the received power \( P_r \) for the current configuration\;
        \nl \If{\( P_r > P_{\text{max}} \)}{
            \nl \( P_{\text{max}} \gets P_r \) \tcp*{Keep new configuration.}
        }\Else{
            \nl Revert columns \( col, col+1 \) to its previous state\;
        }
    }
    \tcc{\textbf{Step 2: Row-Wise Scanning}}
    \nl \For{\texttt{\( row = 1, \dots, N_x \)}}{
        \nl Invert the states of row \( row \) in \( \bm{\Phi} \)\;
        \nl Measure the received power \( P_r \) for the current configuration\;
        \nl \If{\( P_r > P_{\text{max}} \)}{
            \nl \( P_{\text{max}} \gets P_r \) \tcp*{Keep new configuration.}
        }\Else{
            \nl Revert row \( row \) to its previous state\;
        }
    }
\nl \Return{\( \bm{\Phi} \)} \tcp*[l]{Return optimized RIS configuration.}
\end{algorithm}

\section{Indoor and Outdoor Measurement Setups}\label{s:measurement_setup} 
\begin{figure}[t]
\centering
\begin{minipage}{0.47\textwidth}
  \vspace*{2mm}
  \centering
  \includegraphics[width=0.9\textwidth]{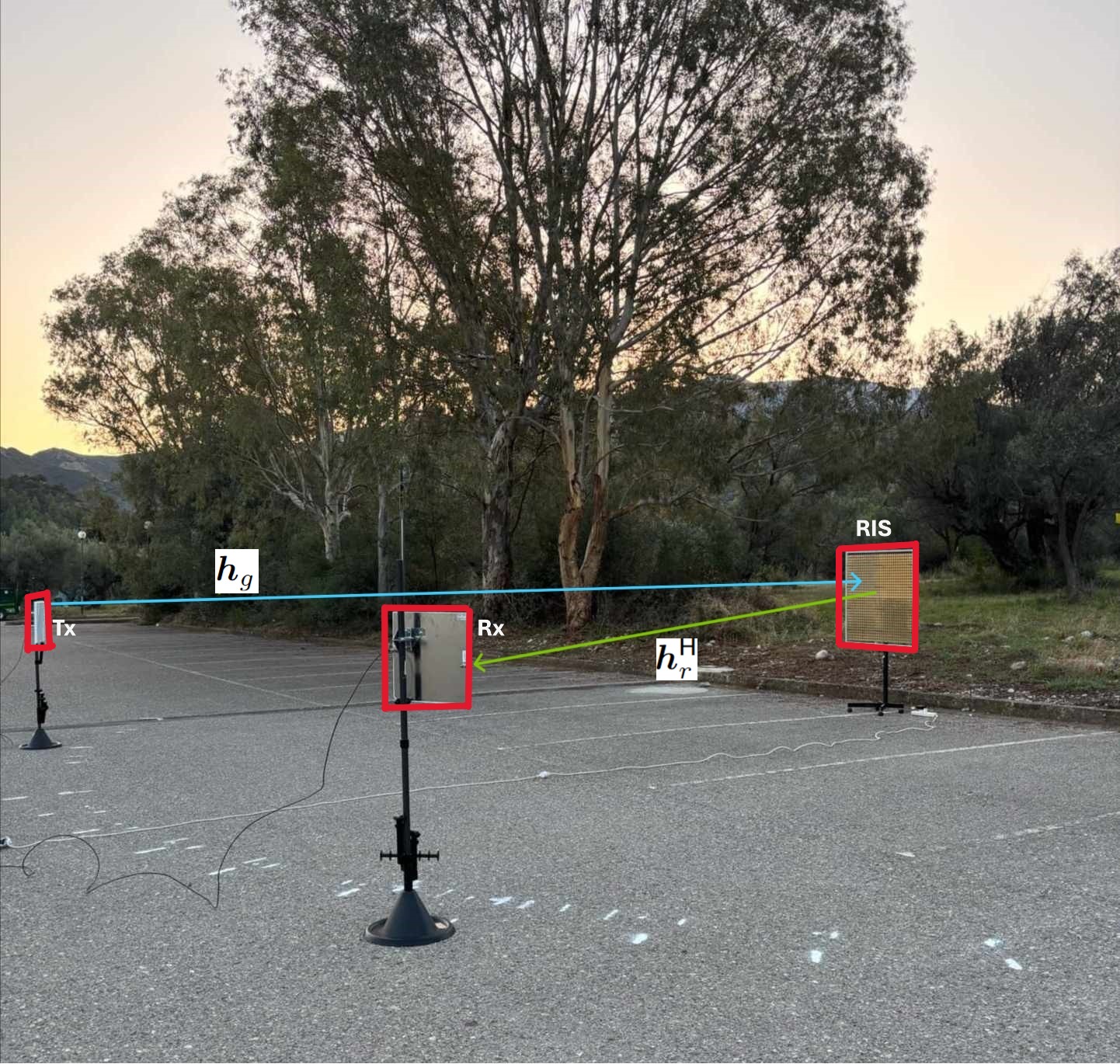}
  \subcaption{Outdoor setup: The Tx-RIS distance is 8.5~m, and the RIS-Rx distance is 8.5~m for all Rx positions, with all nodes positioned at a height of 1.3~m. Rx positions range from $0^\circ$ to $60^\circ$, with a step size of $5^\circ$, 
  and the Tx is fixed at $-15^\circ$. RIS configurations are optimized to steer the beam toward each Rx position.}\label{fig:2a}
\end{minipage}%
\vspace{0.2cm}
\begin{minipage}{0.47\textwidth}
  \centering
  \includegraphics[width=0.9\textwidth]{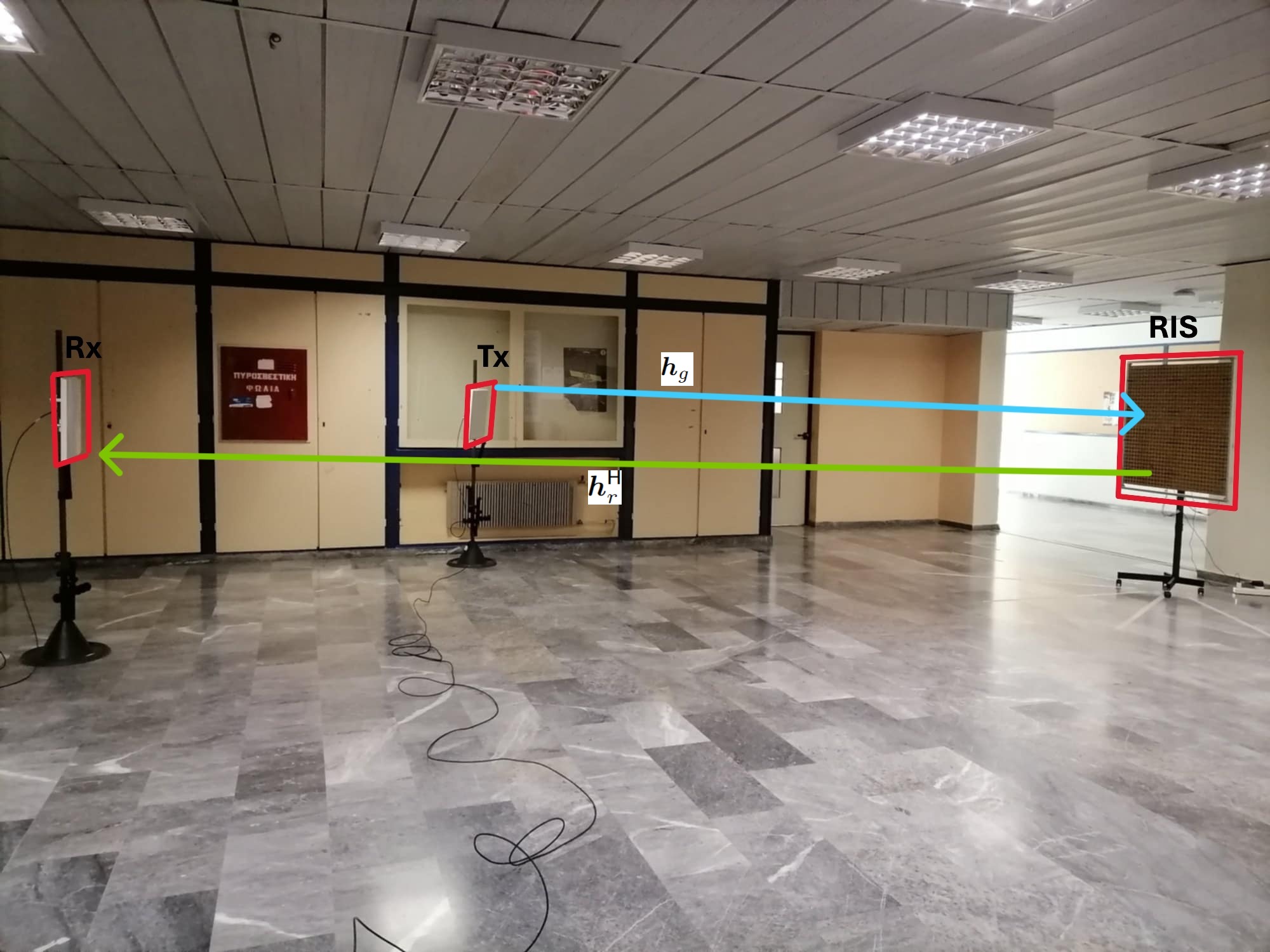}
  \subcaption{Indoor setup: The Tx-RIS distance is 5.5~m, and the RIS-Rx distance is 8.5~m for all Rx positions, with all nodes positioned at a height of 1.3~m. Rx positions range from $0^\circ$ to $45^\circ$, with a step size of $15^\circ$
  and the Tx if fixed at $-15^\circ$. RIS configurations are optimized to steer the beam toward each Rx position.}\label{fig:2c}
\end{minipage}%
\caption{Experimental setups for outdoor and indoor environments.}\label{fig:exp_4images}
\end{figure}

The measurement setups involve both outdoor and indoor experiments (Figs.~\ref{fig:2a} and~\ref{fig:2c}). The RIS, used in both setups, consists of four tiles, each containing a $16{\times}16$ array of unit cells~\cite{rains2023fully}. Each unit cell is equipped with two varactor diodes, each independently driven by two programmable bias voltage levels: $V_1 {=} 11$ V and $V_2 {=} 7.5$ V. The RIS operates around 3.5 GHz and supports dual-polarization (horizontal and vertical). For both transmission and reception, we used flat-panel directional antennas, specifically the PAT3519XP model from ITELITE. These antennas are dual-polarized and have a beamwidth of around $20^\circ$, which reduces the impact of unwanted reflections and interference. The antennas have a working frequency range of 3.5--3.8 GHz. The Vector Network Analyzer (VNA) used in this study is the Copper Mountain S5243, a 2-port device capable of measuring frequencies up to 44 GHz. The VNA was used to measure the S parameters, providing insight into the transmission characteristics between the Tx and the Rx. For accurate measurements, a SOLT (Short-Open-Load-Through) calibration kit T4311 was employed to calibrate the VNA, ensuring precise results.

The outdoor measurement setup (Fig.~\ref{fig:2a}) was conducted in an environment with minimal multipath interference. The Tx and Rx were positioned at a distance of 8.5 meters from the RIS, while all components were placed at a height of 1.3 meters. The measurements were taken with a step size of $5^\circ$ in the azimuth domain. In contrast, the indoor measurement setup (Fig.~\ref{fig:2c}) was conducted in a more challenging environment with strong multipath effects. The nodes were placed at the same height of 1.3 meters, but the distance between the Tx and the RIS was decreased to 5.5 meters. The multipath environment in the indoor setup introduces additional complexity, as signals reflected from walls, floors, and other objects can distort the measurements. The measurement step was $15^\circ$. Furthermore, for communication between the RIS controller (Raspberry Pi) and the VNA, a script-based system was developed. The RIS configuration and VNA were controlled via MATLAB scripts from a laptop, enabling automated measurement and data collection. This setup facilitated efficient testing and ensured that the RIS configuration could be adjusted in real-time during the measurement process, allowing for dynamic investigation.

\section{Experimental Results and Discussion}\label{s:experimental_results}

\subsection{Radiation Pattern of RIS: Codebook-based Beamforming}
In this subsection, we analyze the radiation patterns of the RIS obtained from outdoor experiments (Fig.~\ref{fig:2a}), conducted under a fixed Tx position at $-15^\circ$ and a Rx azimuth range of $0^\circ$ to $60^\circ$ in steps of $5^\circ$ (Fig.~\ref{fig:2a}). At each Rx position, Algorithm~1 was executed to optimize the RIS configuration for maximum received power. For the resulting configurations, we measured the amplitude of the $S_{21}$ signal at all Rx positions, allowing us to extract the RIS radiation patterns. 

\begin{figure*}[t]
    \centering
    \begin{minipage}{0.185\textwidth}
        \centering
        \scalebox{0.24}{
%
%
\begin{tikzpicture}
\useasboundingbox (0,-0.5) rectangle (11.5,6.1);
\begin{axis}[%
width=4.521in,
height=2.26in,
scale only axis,
point meta min=0,
point meta max=1,
axis on top,
xmin=0.5,
xmax=64.5,
xtick={\empty},
xlabel style={font=\color{white!15!black}},
xlabel={\Huge Columns},
y dir=reverse,
ymin=0.5,
ymax=32.5,
ytick={\empty},
ylabel style={font=\color{white!15!black}},
ylabel={\Huge Rows},
axis background/.style={fill=white}
]
\addplot [forget plot] graphics [xmin=0.5, xmax=64.5, ymin=0.5, ymax=32.5] {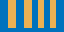};
\end{axis}

\end{tikzpicture}%
}
        \subcaption{RIS Configuration, $0^\circ$}
        \label{fig:ris_conf0}
    \end{minipage}
    \begin{minipage}{0.185\textwidth}
        \centering
        \scalebox{0.24}{
%
%
\begin{tikzpicture}
\useasboundingbox (0,-0.5) rectangle (11.5,6.1);

\begin{axis}[%
width=4.521in,
height=2.26in,
scale only axis,
point meta min=0,
point meta max=1,
axis on top,
xmin=0.5,
xmax=64.5,
xtick={\empty},
xlabel style={font=\color{white!15!black}},
xlabel={\Huge Columns},
y dir=reverse,
ymin=0.5,
ymax=32.5,
ytick={\empty},
ylabel style={font=\color{white!15!black}},
ylabel={\Huge Rows},
axis background/.style={fill=white}
]
\addplot [forget plot] graphics [xmin=0.5, xmax=64.5, ymin=0.5, ymax=32.5] {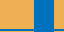};
\end{axis}

\end{tikzpicture}%
}
        \subcaption{RIS Configuration, $15^\circ$}
        \label{fig:ris_conf15}
    \end{minipage}
    \begin{minipage}{0.185\textwidth}
        \centering
        \scalebox{0.24}{
%
%
\begin{tikzpicture}
\useasboundingbox (0,-0.5) rectangle (11.5,6.1);

\begin{axis}[%
width=4.521in,
height=2.26in,
scale only axis,
point meta min=0,
point meta max=1,
axis on top,
xmin=0.5,
xmax=64.5,
xtick={\empty},
xlabel style={font=\color{white!15!black}},
xlabel={\Huge Columns},
y dir=reverse,
ymin=0.5,
ymax=32.5,
ytick={\empty},
ylabel style={font=\color{white!15!black}},
ylabel={\Huge Rows},
axis background/.style={fill=white}
]
\addplot [forget plot] graphics [xmin=0.5, xmax=64.5, ymin=0.5, ymax=32.5] {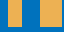};
\end{axis}

\begin{axis}[%
width=5.833in,
height=4.375in,
at={(0in,0in)},
scale only axis,
xmin=0,
xmax=1,
ymin=0,
ymax=1,
axis line style={draw=none},
ticks=none,
axis x line*=bottom,
axis y line*=left
]
\end{axis}
\end{tikzpicture}%
}
        \subcaption{RIS Configuration, $30^\circ$}
        \label{fig:ris_conf30}
    \end{minipage}
    \begin{minipage}{0.185\textwidth}
        \centering
        \scalebox{0.24}{
%
%
\begin{tikzpicture}
\useasboundingbox (0,-0.5) rectangle (11.5,6.1);

\begin{axis}[%
width=4.521in,
height=2.26in,
scale only axis,
point meta min=0,
point meta max=1,
axis on top,
xmin=0.5,
xmax=64.5,
xtick={\empty},
xlabel style={font=\color{white!15!black}},
xlabel={\Huge Columns},
y dir=reverse,
ymin=0.5,
ymax=32.5,
ytick={\empty},
ylabel style={font=\color{white!15!black}},
ylabel={\Huge Rows},
axis background/.style={fill=white}
]
\addplot [forget plot] graphics [xmin=0.5, xmax=64.5, ymin=0.5, ymax=32.5] {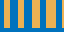};
\end{axis}

\begin{axis}[%
width=5.833in,
height=4.375in,
at={(0in,0in)},
scale only axis,
xmin=0,
xmax=1,
ymin=0,
ymax=1,
axis line style={draw=none},
ticks=none,
axis x line*=bottom,
axis y line*=left
]
\end{axis}
\end{tikzpicture}%
}
        \subcaption{RIS Configuration, $45^\circ$}
        \label{fig:ris_conf45}
    \end{minipage}
    \begin{minipage}{0.185\textwidth}
        \centering
        \scalebox{0.24}{
%
%
\begin{tikzpicture}
\useasboundingbox (0,-0.5) rectangle (11.5,6.1);

\begin{axis}[%
width=4.521in,
height=2.26in,
scale only axis,
point meta min=0,
point meta max=1,
axis on top,
xmin=0.5,
xmax=64.5,
xtick={\empty},
xlabel style={font=\color{white!15!black}},
xlabel={\Huge Columns},
y dir=reverse,
ymin=0.5,
ymax=32.5,
ytick={\empty},
ylabel style={font=\color{white!15!black}},
ylabel={\Huge Rows},
axis background/.style={fill=white}
]
\addplot [forget plot] graphics [xmin=0.5, xmax=64.5, ymin=0.5, ymax=32.5] {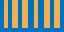};
\end{axis}

\begin{axis}[%
width=5.833in,
height=4.375in,
at={(0in,0in)},
scale only axis,
xmin=0,
xmax=1,
ymin=0,
ymax=1,
axis line style={draw=none},
ticks=none,
axis x line*=bottom,
axis y line*=left
]
\end{axis}
\end{tikzpicture}%
}
        \subcaption{RIS Configuration, $60^\circ$}
        \label{fig:ris_conf60}
    \end{minipage}
    \caption{RIS configurations obtained after executing Algorithm~1 at different Rx azimuth angles. Each configuration corresponds to a $32 \times 64$ RIS, where each element supports both horizontal and vertical polarizations, resulting in double columns. The last rows deviate from the expected column-wise behavior~\cite{bjornson2021optimizing}, likely due to ground bounce effects~\cite{trichopoulos2022design}. This behavior is consistently observed across all the Rx positions of Fig.~\ref{fig:2a}.}
    \label{fig:ris_configs}
\end{figure*}

Bj{\"o}rnson et al.~\cite{bjornson2021optimizing} note that the channel coefficients of RIS elements are approximately equal within each column of the array, when scattering objects are predominantly located in the horizontal plane (limited vertical scattering). In our outdoor setup (Fig.~\ref{fig:2a}), this assumption is reasonable, as the environment lacks tall buildings or other significant vertical scatterers. The area is characterized by sparse trees and open ground, and the Tx, Rx, and RIS were all positioned at the same height of $1.3$ meters, further minimizing vertical reflections. When Algorithm~1 was executed in real time to maximize the received power, the resulting RIS configurations revealed unexpected behavior in the last rows of the array (Fig.~\ref{fig:ris_configs}). Specifically, the lower rows deviated from the expected columnar structure. Given the limited vertical scattering in the outdoor environment, this deviation is likely caused by ground bounce effects, where signals reflected off the ground interact with the direct signal, enhancing or distorting the received power at the Rx. To further investigate this phenomenon, we compared the measured radiation patterns with the theoretical model of Sec.~\ref{s:theoretical_model} under two scenarios: using the measured RIS configuration as input to the model (Case 1), and modifying the last rows by extending the states of the corresponding upper rows within the same columns (Case 2). 

\begin{figure*}[t]
    \centering
    \begin{minipage}{0.185\textwidth}
        \centering
        \scalebox{0.24}{
%
%
\definecolor{mycolor1}{rgb}{0.00000,0.44700,0.74100}%
\definecolor{mycolor2}{rgb}{0.85000,0.32500,0.09800}%
\definecolor{mycolor3}{rgb}{0.92900,0.69400,0.12500}%
\begin{tikzpicture}

\begin{axis}[%
width=4.521in,
height=3.541in,
at={(0.758in,0.506in)},
scale only axis,
xmin=0,
xmax=60,
xlabel style={font=\color{white!15!black}},
xlabel={Angle (degrees)},
ymin=-30,
ymax=0,
ylabel style={font=\color{white!15!black}},
ylabel={Magnitude (dB)},
axis background/.style={fill=white},
title style={font=\bfseries},
xmajorgrids,
ymajorgrids,
legend style={legend cell align=left, align=left, draw=white!15!black}
]
\addplot [color=mycolor1, line width=1.2pt, mark=triangle, mark options={solid, mycolor1}]
  table[row sep=crcr]{%
0	-11.5129348688376\\
1	-13.7050842292267\\
2	-13.5756354850128\\
3	-12.0408566670578\\
4	-11.6339438730106\\
5	-12.9775644874467\\
6	-15.8663377960082\\
7	-17.5955834158791\\
8	-16.5117596140838\\
9	-16.700816214909\\
10	-19.7169500428193\\
11	-19.9624994758085\\
12	-15.0022435367885\\
13	-12.8091258016531\\
14	-13.9418852079273\\
15	-18.9262497449786\\
16	-12.8332786998903\\
17	-6.22732797303474\\
18	-2.55914261947748\\
19	-0.631228758202518\\
20	0\\
21	-0.525657606295546\\
22	-2.25564895609199\\
23	-5.46797142321581\\
24	-10.9360780541338\\
25	-19.1250610027069\\
26	-16.3453438898502\\
27	-13.6272204222029\\
28	-13.9371144140041\\
29	-16.4973481799196\\
30	-20.4720019840877\\
31	-21.5468063933925\\
32	-19.9929298640421\\
33	-19.5337341959903\\
34	-19.5038764608812\\
35	-18.2283863144249\\
36	-16.3542556973449\\
37	-15.2892342440581\\
38	-15.5880196112846\\
39	-17.6844486876801\\
40	-22.4377703441346\\
41	-23.915606287981\\
42	-17.5474642742416\\
43	-13.6685476321782\\
44	-11.5351828584377\\
45	-10.5240429230645\\
46	-10.356031774147\\
47	-10.9004805112211\\
48	-12.1044597909139\\
49	-13.9661506202799\\
50	-16.5146171995214\\
51	-19.7235599975998\\
52	-23.0531427235759\\
53	-24.6091379146533\\
54	-23.8241481214414\\
55	-22.6550774647166\\
56	-21.9317621160654\\
57	-21.6919247828968\\
58	-21.8505122500997\\
59	-22.3426837826813\\
60	-23.1411392308224\\
};
\addlegendentry{Case 1}

\addplot [color=mycolor2, line width=1.2pt, mark=square, mark options={solid, mycolor2}]
  table[row sep=crcr]{%
0	0\\
1	-3.61718194209512\\
2	-6.30775855399673\\
3	-4.08431714739389\\
4	-2.26004814377226\\
5	-2.48411233998906\\
6	-5.07308659816087\\
7	-11.7625485225134\\
8	-24.1229055218373\\
9	-11.4014759081966\\
10	-9.51309529735951\\
11	-12.1704003573212\\
12	-24.8123298476353\\
13	-15.2169935316516\\
14	-9.48234819159022\\
15	-8.29491530945658\\
16	-10.2309542605905\\
17	-17.1083156827063\\
18	-22.7812368936563\\
19	-12.9809103045822\\
20	-10.8830196066196\\
21	-12.6102529969854\\
22	-20.6037618849795\\
23	-19.7114508347825\\
24	-11.571741756848\\
25	-9.07312559765606\\
26	-9.42684574108975\\
27	-12.8417004099238\\
28	-24.1721062885048\\
29	-18.1976350861399\\
30	-12.0537489407605\\
31	-10.7833599241706\\
32	-12.9315977842362\\
33	-22.3882558768902\\
34	-17.0106993983827\\
35	-9.06911597886639\\
36	-5.66096040007726\\
37	-4.28070781418486\\
38	-4.35414709706529\\
39	-5.71057255953222\\
40	-7.92163657409978\\
41	-8.54906052083734\\
42	-6.06111590014163\\
43	-3.41092665690334\\
44	-1.65617049529254\\
45	-0.747105026630287\\
46	-0.544171426263375\\
47	-0.94532573983777\\
48	-1.87798123808841\\
49	-3.27255832035343\\
50	-5.01723185988101\\
51	-6.88115115312706\\
52	-8.46413940915802\\
53	-9.40304457553604\\
54	-9.76429985396491\\
55	-9.93606894308226\\
56	-10.2299228740659\\
57	-10.7829557584668\\
58	-11.6290868061431\\
59	-12.7578161812497\\
60	-14.1392620024048\\
};
\addlegendentry{Case 2}

\addplot [color=mycolor3, dashed, line width=1.5pt, mark size=4.0pt, mark=o, mark options={solid, mycolor3}]
  table[row sep=crcr]{%
0	0\\
5	-9.83385\\
10	-28.10913\\
15	-18.83296\\
20	-20.11346\\
25	-13.50132\\
30	-6.51126\\
35	-6.59193999999999\\
40	-6.69179\\
45	-3.09581\\
50	-9.64471999999999\\
55	-8.58743\\
60	-20.44139\\
};
\addlegendentry{Real Measurements}

\end{axis}

\begin{axis}[%
width=5.833in,
height=4.375in,
at={(0in,0in)},
scale only axis,
xmin=0,
xmax=1,
ymin=0,
ymax=1,
axis line style={draw=none},
ticks=none,
axis x line*=bottom,
axis y line*=left
]
\end{axis}
\end{tikzpicture}%
}
        \subcaption{RIS Configuration, $0^\circ$}
        \label{fig:ris_val0}
    \end{minipage}
    \begin{minipage}{0.185\textwidth}
        \centering
        \scalebox{0.24}{
%
%
\definecolor{mycolor1}{rgb}{0.00000,0.44700,0.74100}%
\definecolor{mycolor2}{rgb}{0.85000,0.32500,0.09800}%
\definecolor{mycolor3}{rgb}{0.92900,0.69400,0.12500}%
\begin{tikzpicture}

\begin{axis}[%
width=4.521in,
height=3.541in,
at={(0.758in,0.506in)},
scale only axis,
xmin=0,
xmax=60,
xlabel style={font=\color{white!15!black}},
xlabel={Angle (degrees)},
ymin=-45,
ymax=0,
ylabel style={font=\color{white!15!black}},
ylabel={Magnitude (dB)},
axis background/.style={fill=white},
title style={font=\bfseries},
xmajorgrids,
ymajorgrids,
legend style={legend cell align=left, align=left, draw=white!15!black}
]
\addplot [color=mycolor1, line width=1.2pt, mark=triangle, mark options={solid, mycolor1}]
  table[row sep=crcr]{%
0	-14.6786445593638\\
1	-17.0800408733408\\
2	-14.8221117841331\\
3	-13.5056047891839\\
4	-15.7856436102923\\
5	-29.5169701153514\\
6	-15.9040707197746\\
7	-9.40503594088598\\
8	-6.93284718905295\\
9	-6.77561145313041\\
10	-8.35403798275549\\
11	-9.40938717767622\\
12	-7.01557171159772\\
13	-4.50856116296803\\
14	-3.49113732709947\\
15	-3.83236348857269\\
16	-4.59908566402909\\
17	-3.87428667854773\\
18	-1.87680686276372\\
19	-0.372671046466081\\
20	0\\
21	-0.892699298254037\\
22	-3.09602062734172\\
23	-6.01945913050772\\
24	-6.64963169336834\\
25	-4.82937441625639\\
26	-3.67830114366071\\
27	-3.90136616491346\\
28	-5.49019095298853\\
29	-8.15968422143313\\
30	-10.1649551842161\\
31	-9.47124381848423\\
32	-8.35807711694064\\
33	-8.44069093363919\\
34	-10.0983986446108\\
35	-13.9183630270922\\
36	-22.690098202867\\
37	-26.8884590976916\\
38	-17.8890155948305\\
39	-15.5750266534219\\
40	-15.7167231911708\\
41	-17.4382163622038\\
42	-19.3545874196389\\
43	-18.6468032239888\\
44	-16.4785826708205\\
45	-15.0035579496104\\
46	-14.5455228123771\\
47	-15.0332126212438\\
48	-16.4219941456527\\
49	-18.7575172475617\\
50	-22.2283723048995\\
51	-27.337995423458\\
52	-35.5774703149042\\
53	-42.2588259608004\\
54	-36.3544858784211\\
55	-34.0019158308331\\
56	-32.3725032130566\\
57	-30.4442894575694\\
58	-28.5674418649826\\
59	-27.1597522715976\\
60	-26.3365627168834\\
};
\addlegendentry{Case 1}

\addplot [color=mycolor2, line width=1.2pt, mark=square, mark options={solid, mycolor2}]
  table[row sep=crcr]{%
0	-11.6192839004329\\
1	-13.296753802071\\
2	-12.358943880634\\
3	-11.9407319361271\\
4	-14.8366253171932\\
5	-31.4335197884916\\
6	-13.4867683306669\\
7	-7.35332925400504\\
8	-4.81062451847774\\
9	-4.25584248268624\\
10	-4.87051615838661\\
11	-4.92688111611302\\
12	-3.11083845103845\\
13	-1.07327373125371\\
14	0\\
15	-0.0795775819830951\\
16	-1.20274085762289\\
17	-2.87425948114361\\
18	-3.87692904398538\\
19	-3.6699421474094\\
20	-3.39697984751712\\
21	-3.72908721758092\\
22	-4.1337299181391\\
23	-3.47326511809111\\
24	-2.00041443853403\\
25	-0.864923095827358\\
26	-0.597350407232916\\
27	-1.31098076855157\\
28	-2.94289087447152\\
29	-5.03319920149285\\
30	-6.27090317931094\\
31	-6.02489368202804\\
32	-5.64862211995583\\
33	-6.17053697587328\\
34	-8.02761031739265\\
35	-11.8196325140833\\
36	-19.8323641158631\\
37	-27.3346272917987\\
38	-17.1076569192978\\
39	-14.4727395003843\\
40	-14.2614450000246\\
41	-15.2375839462344\\
42	-15.8975274502406\\
43	-14.9043805370014\\
44	-13.262392466174\\
45	-12.1202409338504\\
46	-11.7511724347751\\
47	-12.1476408691622\\
48	-13.2757960159723\\
49	-15.1279967574309\\
50	-17.7217750986647\\
51	-21.0384713716032\\
52	-24.7078840076072\\
53	-27.2621392903744\\
54	-27.5734737197778\\
55	-27.0632556535158\\
56	-26.6904425836637\\
57	-26.4637470033469\\
58	-26.2336846752061\\
59	-25.974297356617\\
60	-25.7641067915393\\
};
\addlegendentry{Case 2}

\addplot [color=mycolor3, dashed, line width=1.5pt, mark size=4.0pt, mark=o, mark options={solid, mycolor3}]
  table[row sep=crcr]{%
0	-14.0981\\
5	-16.30763\\
10	-13.7896\\
15	0\\
20	-6.97848\\
25	-5.51293\\
30	-5.40864999999999\\
35	-18.30925\\
40	-27.21271\\
45	-9.65184\\
50	-21.56723\\
55	-22.70132\\
60	-16.94833\\
};
\addlegendentry{Real Measurements}

\end{axis}

\begin{axis}[%
width=5.833in,
height=4.375in,
at={(0in,0in)},
scale only axis,
xmin=0,
xmax=1,
ymin=0,
ymax=1,
axis line style={draw=none},
ticks=none,
axis x line*=bottom,
axis y line*=left
]
\end{axis}
\end{tikzpicture}%
}
        \subcaption{RIS Configuration, $15^\circ$}
        \label{fig:ris_val15}
    \end{minipage}
    \begin{minipage}{0.185\textwidth}
        \centering
        \scalebox{0.24}{
%
%
\definecolor{mycolor1}{rgb}{0.00000,0.44700,0.74100}%
\definecolor{mycolor2}{rgb}{0.85000,0.32500,0.09800}%
\definecolor{mycolor3}{rgb}{0.92900,0.69400,0.12500}%
\begin{tikzpicture}

\begin{axis}[%
width=4.521in,
height=3.541in,
at={(0.758in,0.506in)},
scale only axis,
xmin=0,
xmax=60,
xlabel style={font=\color{white!15!black}},
xlabel={Angle (degrees)},
ymin=-50,
ymax=0,
ylabel style={font=\color{white!15!black}},
ylabel={Magnitude (dB)},
axis background/.style={fill=white},
title style={font=\bfseries},
xmajorgrids,
ymajorgrids,
legend style={legend cell align=left, align=left, draw=white!15!black}
]
\addplot [color=mycolor1, line width=1.2pt, mark=triangle, mark options={solid, mycolor1}]
  table[row sep=crcr]{%
0	-17.7940355684854\\
1	-24.8654847636706\\
2	-27.0415588051027\\
3	-18.3690621348448\\
4	-16.2425604075084\\
5	-17.8110183609841\\
6	-24.6902647585152\\
7	-22.1241577003038\\
8	-15.240427890282\\
9	-12.4217459165332\\
10	-11.4217475083697\\
11	-11.022218120302\\
12	-10.5092189880369\\
13	-10.3589333142213\\
14	-11.8526018919031\\
15	-17.5642403397297\\
16	-17.4388124538197\\
17	-8.11068712431223\\
18	-3.52708031953844\\
19	-1.06257227213744\\
20	0\\
21	-0.110337394548267\\
22	-1.38931498464537\\
23	-4.0588189160162\\
24	-8.81571178117845\\
25	-17.3411064361609\\
26	-15.7203634950053\\
27	-11.1617988296857\\
28	-9.82885069937697\\
29	-10.2460622347398\\
30	-11.6529094704845\\
31	-13.1205657730966\\
32	-13.7757949981142\\
33	-14.006807910519\\
34	-14.789347235054\\
35	-16.7499169788315\\
36	-20.4362027334653\\
37	-26.2059852441214\\
38	-26.7460532691807\\
39	-22.9739971871815\\
40	-21.364130833956\\
41	-21.5097884001658\\
42	-23.122489367314\\
43	-26.3299137274895\\
44	-32.0369880577097\\
45	-46.6070941634486\\
46	-38.2416980512857\\
47	-32.4368496167291\\
48	-29.7052958865212\\
49	-27.8931035308398\\
50	-26.5275663736342\\
51	-25.5678802310666\\
52	-25.0580858406368\\
53	-25.0240634879658\\
54	-25.467094962961\\
55	-26.3757589133858\\
56	-27.7328903361272\\
57	-29.5143073800928\\
58	-31.6776310771356\\
59	-34.1362829794769\\
60	-36.7140919648661\\
};
\addlegendentry{Case 1}

\addplot [color=mycolor2, line width=1.2pt, mark=square, mark options={solid, mycolor2}]
  table[row sep=crcr]{%
0	-9.73516826797754\\
1	-11.4236082251063\\
2	-15.3757750542248\\
3	-14.9036208680596\\
4	-10.8968557130599\\
5	-9.11128485301003\\
6	-8.87957394654747\\
7	-8.01170149262224\\
8	-5.20894224840857\\
9	-2.40160240808873\\
10	-0.631637342016823\\
11	0\\
12	-0.47105172111776\\
13	-2.0272805732443\\
14	-4.55512480151047\\
15	-7.26708127045966\\
16	-8.43083212972437\\
17	-8.48202901634885\\
18	-9.30326265028334\\
19	-11.5034724720747\\
20	-13.4896003945048\\
21	-11.9331521139202\\
22	-9.753356503448\\
23	-8.87917566843652\\
24	-8.92323971109674\\
25	-8.35697763814883\\
26	-6.20499387496982\\
27	-3.77391484854353\\
28	-2.04196832071218\\
29	-1.19794002050613\\
30	-1.23073863868959\\
31	-2.12255311225512\\
32	-3.86447176692133\\
33	-6.35251219651573\\
34	-9.00020930197668\\
35	-10.4946711456623\\
36	-10.7606987382784\\
37	-11.270884068418\\
38	-12.9354624205865\\
39	-16.0474719354224\\
40	-18.7313303268885\\
41	-16.6208634041064\\
42	-13.9170845511137\\
43	-12.6329132496385\\
44	-12.6799774284245\\
45	-13.9700980886243\\
46	-16.5111492086205\\
47	-19.6395549912566\\
48	-19.7752853511462\\
49	-17.158292141968\\
50	-15.13586365843\\
51	-14.1512814805332\\
52	-14.0375709783467\\
53	-14.637415325167\\
54	-15.7961695458648\\
55	-17.2455089743144\\
56	-18.4573115858446\\
57	-18.8503275942779\\
58	-18.4844862187641\\
59	-17.9326286839047\\
60	-17.6054051301608\\
};
\addlegendentry{Case 2}

\addplot [color=mycolor3, dashed, line width=1.5pt, mark size=4.0pt, mark=o, mark options={solid, mycolor3}]
  table[row sep=crcr]{%
0	-10.48016\\
5	-8.56865999999999\\
10	-7.24165\\
15	-8.69526\\
20	-14.31\\
25	-23.88117\\
30	0\\
35	-22.79826\\
40	-24.16076\\
45	-15.24653\\
50	-23.59478\\
55	-19.09318\\
60	-9.30128999999999\\
};
\addlegendentry{Real Measurements}

\end{axis}

\begin{axis}[%
width=5.833in,
height=4.375in,
at={(0in,0in)},
scale only axis,
xmin=0,
xmax=1,
ymin=0,
ymax=1,
axis line style={draw=none},
ticks=none,
axis x line*=bottom,
axis y line*=left
]
\end{axis}
\end{tikzpicture}%
}
        \subcaption{RIS Configuration, $30^\circ$}
        \label{fig:ris_val30}
    \end{minipage}
    \begin{minipage}{0.185\textwidth}
        \centering
        \scalebox{0.24}{
%
%
\definecolor{mycolor1}{rgb}{0.00000,0.44700,0.74100}%
\definecolor{mycolor2}{rgb}{0.85000,0.32500,0.09800}%
\definecolor{mycolor3}{rgb}{0.92900,0.69400,0.12500}%
\begin{tikzpicture}

\begin{axis}[%
width=4.521in,
height=3.541in,
at={(0.758in,0.506in)},
scale only axis,
xmin=0,
xmax=60,
xlabel style={font=\color{white!15!black}},
xlabel={Angle (degrees)},
ymin=-50,
ymax=0,
ylabel style={font=\color{white!15!black}},
ylabel={Magnitude (dB)},
axis background/.style={fill=white},
title style={font=\bfseries},
xmajorgrids,
ymajorgrids,
legend style={legend cell align=left, align=left, draw=white!15!black}
]
\addplot [color=mycolor1, line width=1.2pt, mark=triangle, mark options={solid, mycolor1}]
  table[row sep=crcr]{%
0	-7.38627530305651\\
1	-10.7227815011013\\
2	-17.5019522148272\\
3	-29.02114441177\\
4	-21.0000911566035\\
5	-20.5380614191636\\
6	-24.7330295001669\\
7	-23.5145081882792\\
8	-19.5763086532215\\
9	-19.7654673357314\\
10	-24.898521549952\\
11	-22.8626534073272\\
12	-16.769823726512\\
13	-15.4037443007262\\
14	-19.4816900723444\\
15	-24.7943589805062\\
16	-10.578216071032\\
17	-4.95706714953497\\
18	-1.87802127536114\\
19	-0.333420097852631\\
20	0\\
21	-0.80718096875006\\
22	-2.8826374529654\\
23	-6.70752725580306\\
24	-14.1120472686386\\
25	-29.5526834010377\\
26	-14.6644971281699\\
27	-11.9359125098044\\
28	-12.3085743043788\\
29	-15.1319317797964\\
30	-21.5833990345836\\
31	-27.1922508609845\\
32	-19.7663689800197\\
33	-17.109614122701\\
34	-16.9923768026088\\
35	-18.8354433861367\\
36	-22.658322943291\\
37	-24.8619215188787\\
38	-20.5374310500864\\
39	-16.8699336588488\\
40	-14.5499015078944\\
41	-13.0942177657324\\
42	-12.2240437466192\\
43	-11.7981901089804\\
44	-11.7501612652247\\
45	-12.0543013201944\\
46	-12.7076044978569\\
47	-13.7202511487152\\
48	-15.109797069213\\
49	-16.8920989362379\\
50	-19.0532484625294\\
51	-21.4633615610408\\
52	-23.6955858677253\\
53	-25.0489026536013\\
54	-25.3706705464504\\
55	-25.3202857442556\\
56	-25.4637721261425\\
57	-25.997036979652\\
58	-26.8920729667458\\
59	-27.9458579763688\\
60	-28.7716854387964\\
};
\addlegendentry{Case 1}

\addplot [color=mycolor2, line width=1.2pt, mark=square, mark options={solid, mycolor2}]
  table[row sep=crcr]{%
0	0\\
1	-2.02789022329744\\
2	-5.60404570651727\\
3	-11.1056885409837\\
4	-15.6396008541408\\
5	-14.174954597591\\
6	-13.8966124958242\\
7	-16.3463646120122\\
8	-21.7262979882597\\
9	-22.7573878840743\\
10	-18.7405447062638\\
11	-17.0562946419164\\
12	-16.7011369200612\\
13	-16.3545900674608\\
14	-15.4439095819127\\
15	-14.564199890685\\
16	-14.3786969432511\\
17	-15.2564732646953\\
18	-17.6331390887217\\
19	-22.82936025942\\
20	-48.1875267876959\\
21	-23.9273326423491\\
22	-18.3492841788161\\
23	-15.8776414290346\\
24	-14.9278074706159\\
25	-15.0105445147241\\
26	-15.8070894060117\\
27	-16.8471523559926\\
28	-17.5259534211015\\
29	-17.8183483837649\\
30	-18.508251514734\\
31	-20.51457462968\\
32	-24.2753859897679\\
33	-23.9474642168593\\
34	-19.1965419999276\\
35	-16.3938779825637\\
36	-15.5656308201291\\
37	-16.4732831411678\\
38	-17.768201931446\\
39	-14.8689333954203\\
40	-10.3055748485942\\
41	-6.86358094649352\\
42	-4.5303333025673\\
43	-3.07294043952391\\
44	-2.3424689592481\\
45	-2.25944886836809\\
46	-2.79441282708032\\
47	-3.96144930867318\\
48	-5.82313810190711\\
49	-8.50229560060824\\
50	-12.145506539294\\
51	-16.3048777532599\\
52	-17.7273549694122\\
53	-16.0741461238658\\
54	-14.8425833898965\\
55	-14.7024316090321\\
56	-15.6136299626227\\
57	-17.6190241277302\\
58	-20.9426136273968\\
59	-25.4732790385581\\
60	-26.2049244949485\\
};
\addlegendentry{Case 2}

\addplot [color=mycolor3, dashed, line width=1.5pt, mark size=4.0pt, mark=o, mark options={solid, mycolor3}]
  table[row sep=crcr]{%
0	-5.3293\\
5	-6.95337\\
10	-11.39936\\
15	-14.53505\\
20	-24.18032\\
25	-12.80109\\
30	-21.61014\\
35	-16.92446\\
40	-4.4235\\
45	0\\
50	-9.73663\\
55	-12.59637\\
60	-46.17454\\
};
\addlegendentry{Real Measurements}

\end{axis}

\begin{axis}[%
width=5.833in,
height=4.375in,
at={(0in,0in)},
scale only axis,
xmin=0,
xmax=1,
ymin=0,
ymax=1,
axis line style={draw=none},
ticks=none,
axis x line*=bottom,
axis y line*=left
]
\end{axis}
\end{tikzpicture}%
}
        \subcaption{RIS Configuration, $45^\circ$}
        \label{fig:ris_val45}
    \end{minipage}
    \begin{minipage}{0.185\textwidth}
        \centering
        \scalebox{0.24}{
%
%
\definecolor{mycolor1}{rgb}{0.00000,0.44700,0.74100}%
\definecolor{mycolor2}{rgb}{0.85000,0.32500,0.09800}%
\definecolor{mycolor3}{rgb}{0.92900,0.69400,0.12500}%
\begin{tikzpicture}

\begin{axis}[%
width=4.521in,
height=3.541in,
at={(0.758in,0.506in)},
scale only axis,
xmin=0,
xmax=60,
xlabel style={font=\color{white!15!black}},
xlabel={Angle (degrees)},
ymin=-70,
ymax=0,
ylabel style={font=\color{white!15!black}},
ylabel={Magnitude (dB)},
axis background/.style={fill=white},
title style={font=\bfseries},
xmajorgrids,
ymajorgrids,
legend style={at={(0.664,0.325)}, anchor=south west, legend cell align=left, align=left, draw=white!15!black}
]
\addplot [color=mycolor1, line width=1.2pt, mark=triangle, mark options={solid, mycolor1}]
  table[row sep=crcr]{%
0	-25.5203457592778\\
1	-23.183618367366\\
2	-20.7668994643841\\
3	-19.7806055700259\\
4	-20.9011543325049\\
5	-24.6387292563742\\
6	-25.4425040343947\\
7	-21.0621120395934\\
8	-19.4758675266653\\
9	-20.5166277919642\\
10	-19.6795497065858\\
11	-15.0207208735997\\
12	-11.949336531828\\
13	-11.2984708391168\\
14	-13.9159198401671\\
15	-29.8442511031028\\
16	-13.2085336148329\\
17	-5.99534716625482\\
18	-2.38528097257403\\
19	-0.551505869010235\\
20	0\\
21	-0.587397892058291\\
22	-2.36701234591677\\
23	-5.64325392325986\\
24	-11.4082667841874\\
25	-25.9269882293028\\
26	-19.9987379990393\\
27	-15.4570995152843\\
28	-15.6085289993516\\
29	-18.3295500039001\\
30	-20.864947074112\\
31	-18.7435430936635\\
32	-16.6914930762859\\
33	-16.4889992313291\\
34	-18.1639496724935\\
35	-22.0190405621759\\
36	-27.5621675469527\\
37	-26.0182125347212\\
38	-23.0319649886917\\
39	-22.4598688953885\\
40	-23.8755853501399\\
41	-26.6499893459564\\
42	-27.5422335052737\\
43	-25.1792222045868\\
44	-23.3574111597398\\
45	-22.96350055915\\
46	-23.9995626718365\\
47	-26.5018055938734\\
48	-29.2585218133303\\
49	-27.1234849225123\\
50	-23.0077203437503\\
51	-19.850420693136\\
52	-17.5349345227492\\
53	-15.7963188938229\\
54	-14.4835147436439\\
55	-13.5215939548102\\
56	-12.8760624776877\\
57	-12.532025095722\\
58	-12.4838201634369\\
59	-12.7309359741697\\
60	-13.2773527213823\\
};
\addlegendentry{Case 1}

\addplot [color=mycolor2, line width=1.2pt, mark=square, mark options={solid, mycolor2}]
  table[row sep=crcr]{%
0	-12.3668794253823\\
1	-12.1180947834066\\
2	-16.0060402779874\\
3	-35.0024402362127\\
4	-17.8825089110535\\
5	-14.0135959999599\\
6	-15.0871876462829\\
7	-19.6083797919764\\
8	-14.1985513232049\\
9	-8.98972815588516\\
10	-6.67887911461571\\
11	-6.35517575534485\\
12	-7.76477020954745\\
13	-11.0761231471171\\
14	-16.9061666223806\\
15	-23.2377333549026\\
16	-21.5019133291553\\
17	-21.5719754230241\\
18	-25.2993632398403\\
19	-33.8116849609858\\
20	-62.6081082135945\\
21	-35.4123726799376\\
22	-26.350864618765\\
23	-22.2794107155319\\
24	-21.6805704462616\\
25	-23.7922781375901\\
26	-20.1171601626603\\
27	-13.9181583894865\\
28	-10.0787696669255\\
29	-8.03693554549653\\
30	-7.43147386536054\\
31	-8.18702544604027\\
32	-10.4862374988419\\
33	-14.8923818710259\\
34	-20.9556607917509\\
35	-18.7659202479479\\
36	-16.0544313707786\\
37	-16.2745564038064\\
38	-19.7991362088531\\
39	-33.1861787809855\\
40	-22.9388044653328\\
41	-16.7899892751314\\
42	-14.6697127377439\\
43	-14.7357177987329\\
44	-16.8404191929102\\
45	-21.7355561590666\\
46	-25.9153486451269\\
47	-19.840220488153\\
48	-16.2038444534699\\
49	-14.2831949508353\\
50	-12.7221769675957\\
51	-10.6177057606074\\
52	-8.05879750660637\\
53	-5.60975080754059\\
54	-3.57852737964693\\
55	-2.03099454406419\\
56	-0.944996240948548\\
57	-0.280224467351651\\
58	0\\
59	-0.076867884641274\\
60	-0.49379692649061\\
};
\addlegendentry{Case 2}

\addplot [color=mycolor3, dashed, line width=1.5pt, mark size=4.0pt, mark=o, mark options={solid, mycolor3}]
  table[row sep=crcr]{%
0	-11.531\\
5	-16.1673\\
10	-16.41876\\
15	-20.17275\\
20	-23.38745\\
25	-13.47651\\
30	-16.29332\\
35	-19.83092\\
40	-11.36462\\
45	-18.27192\\
50	-18.66712\\
55	-3.5953\\
60	0\\
};
\addlegendentry{Real Measurements}

\end{axis}

\begin{axis}[%
width=5.833in,
height=4.375in,
at={(0in,0in)},
scale only axis,
xmin=0,
xmax=1,
ymin=0,
ymax=1,
axis line style={draw=none},
ticks=none,
axis x line*=bottom,
axis y line*=left
]
\end{axis}
\end{tikzpicture}%
}
        \subcaption{RIS Configuration, $60^\circ$}
        \label{fig:ris_val60}
    \end{minipage}
    \caption{Validation of RIS radiation patterns measured in the outdoor setup (Fig.~\ref{fig:2a}) at Rx positions ranging from $0^\circ$ to $60^\circ$, with a step size of $5^\circ$ (3.55 GHz). Case 1 corresponds to the theoretical radiation pattern model~\cite{ramezani2024broad,10256051} derived using the RIS configurations generated by Algorithm 1, while Case 2 represents the theoretical model by modifying the last rows of the RIS configuration matrix, specifically extending the states of the corresponding upper rows within the same columns. The gains are normalized, with the maximum value set to $0\,\mathrm{dB}$. It is noted that the theoretical patterns for Case 1 and Case 2 are plotted with a resolution of $1^\circ$ while the measured patterns are taken at $5^\circ$ intervals. The maximum gain across all Rx positions is observed at the location where beam optimization via Algorithm 1 was performed.}
    \label{fig:validation}
\end{figure*}
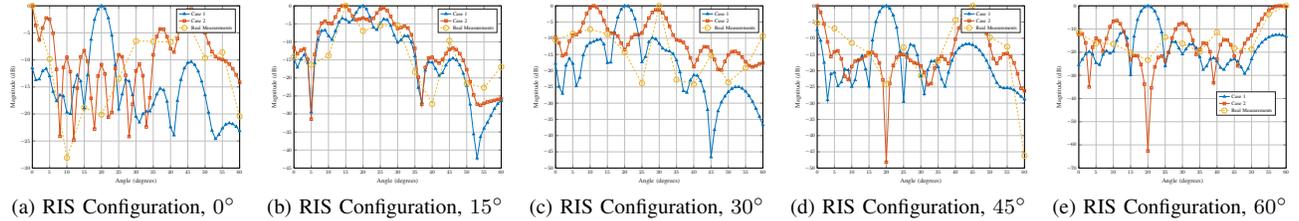

Fig.~\ref{fig:validation} compares the measured radiation patterns with the theoretical patterns derived from Case 1 and Case 2. 
The measurements strongly validate the theoretical radiation pattern model, with Case 2 providing enhanced alignment by accounting for ground-bounce effects in the last rows. Furthermore, each radiation pattern consistently exhibits a pronounced main lobe directed toward the desired Rx angle, even for angles near the edge of the RIS scanning range (e.g., $60^\circ$), demonstrating the beamforming capability of the RIS. This highlights that the proposed codebook is well-suited for beam sweeping and is further utilized in the subsequent subsections.

\subsection{Impact of Frequency Selectivity on RIS Beamforming} 
\begin{figure}[t]
\centering
\begin{minipage}{0.24\textwidth}
    \centering
    \scalebox{0.29}{
%
%
\definecolor{mycolor1}{rgb}{0.00000,0.44700,0.74100}%
\definecolor{mycolor2}{rgb}{0.85000,0.32500,0.09800}%
\definecolor{mycolor3}{rgb}{0.92900,0.69400,0.12500}%
\begin{tikzpicture}

\begin{axis}[%
width=4.844in,
height=3.396in,
at={(0.812in,0.458in)},
scale only axis,
xmin=0,
xmax=60,
xlabel style={font=\color{white!15!black}},
xlabel={Angle (degrees)},
ymin=-90,
ymax=-50,
ylabel style={font=\color{white!15!black}},
ylabel={Magnitude (dB)},
axis background/.style={fill=white},
title style={font=\bfseries},
axis x line*=bottom,
axis y line*=left,
xmajorgrids,
ymajorgrids,
legend style={legend cell align=left, align=left, draw=white!15!black}
]
\addplot [color=mycolor1, line width=1.5pt, mark=o, mark options={solid, mycolor1}]
  table[row sep=crcr]{%
0	-60.12\\
5	-72.06488\\
10	-72.0265\\
15	-76.54573\\
20	-73.92595\\
25	-72.11463\\
30	-70.84872\\
35	-71.36137\\
40	-66.66101\\
45	-67.20325\\
50	-66.59303\\
55	-71.88374\\
60	-74.61409\\
};
\addlegendentry{3.50 GHz}

\addplot [color=mycolor2, dashed, line width=1.5pt, mark size=4.0pt, mark=asterisk, mark options={solid, mycolor2}]
  table[row sep=crcr]{%
0	-56.87895\\
5	-70.73302\\
10	-68.75427\\
15	-69.49359\\
20	-64.73093\\
25	-72.60362\\
30	-66.88623\\
35	-64.37363\\
40	-64.31977\\
45	-63.09956\\
50	-63.89492\\
55	-68.32399\\
60	-70.94555\\
};
\addlegendentry{3.55 GHz}

\addplot [color=mycolor3, dotted, line width=1.5pt, mark size=6.0pt, mark=diamond, mark options={solid, mycolor3}]
  table[row sep=crcr]{%
0	-56.25191\\
5	-72.12994\\
10	-66.09411\\
15	-64.7939\\
20	-63.22687\\
25	-70.17367\\
30	-66.40355\\
35	-64.49361\\
40	-63.65503\\
45	-61.24127\\
50	-62.15111\\
55	-66.85991\\
60	-74.83644\\
};
\addlegendentry{3.60 GHz}

\end{axis}
\end{tikzpicture}%
}
    \subcaption{RIS Configuration at $0^\circ$}\label{fig:freqsel0}
\end{minipage}%
\hfill
\begin{minipage}{0.24\textwidth}
    \centering
    \scalebox{0.29}{
%
%
\definecolor{mycolor1}{rgb}{0.00000,0.44700,0.74100}%
\definecolor{mycolor2}{rgb}{0.85000,0.32500,0.09800}%
\definecolor{mycolor3}{rgb}{0.92900,0.69400,0.12500}%
\begin{tikzpicture}

\begin{axis}[%
width=4.844in,
height=3.396in,
at={(0.812in,0.458in)},
scale only axis,
xmin=0,
xmax=60,
xlabel style={font=\color{white!15!black}},
xlabel={Angle (degrees)},
ymin=-90,
ymax=-50,
ylabel style={font=\color{white!15!black}},
ylabel={Magnitude (dB)},
axis background/.style={fill=white},
title style={font=\bfseries},
axis x line*=bottom,
axis y line*=left,
xmajorgrids,
ymajorgrids,
legend style={legend cell align=left, align=left, draw=white!15!black}
]
\addplot [color=mycolor1, line width=1.5pt, mark=o, mark options={solid, mycolor1}]
  table[row sep=crcr]{%
0	-65.02938\\
5	-67.92269\\
10	-65.64189\\
15	-66.28907\\
20	-68.47385\\
25	-77.32084\\
30	-60.38505\\
35	-72.44364\\
40	-75.05635\\
45	-85.20625\\
50	-81.35133\\
55	-71.66863\\
60	-76.00131\\
};
\addlegendentry{3.50 GHz}

\addplot [color=mycolor2, dashed, line width=1.5pt, mark size=4.0pt, mark=asterisk, mark options={solid, mycolor2}]
  table[row sep=crcr]{%
0	-63.39021\\
5	-67.2532\\
10	-62.14983\\
15	-60.80414\\
20	-75.7338\\
25	-73.21961\\
30	-56.40607\\
35	-74.7845\\
40	-78.35597\\
45	-79.3804\\
50	-72.04745\\
55	-69.43168\\
60	-75.70787\\
};
\addlegendentry{3.55 GHz}

\addplot [color=mycolor3, dotted, line width=1.5pt, mark size=6.0pt, mark=diamond, mark options={solid, mycolor3}]
  table[row sep=crcr]{%
0	-65.38361\\
5	-69.19158\\
10	-60.99902\\
15	-60.90404\\
20	-64.38181\\
25	-71.69624\\
30	-57.00103\\
35	-80.06198\\
40	-80.32808\\
45	-80.03744\\
50	-70.21457\\
55	-72.0845\\
60	-78.05464\\
};
\addlegendentry{3.60 GHz}

\end{axis}
\end{tikzpicture}%
}
    \subcaption{RIS Configuration at $30^\circ$}\label{fig:freq_sel30}
\end{minipage}%
\caption{Comparison of measured radiation patterns across frequencies. The results demonstrate that the angle direction of the main lobe is preserved across the frequency range, showcasing the effectiveness of the proposed codebook for beam sweeping in wideband applications. Variations are observed in the side lobes. Similar results in all Rx positions.}
\label{fig:freq_selectivity}
\end{figure}
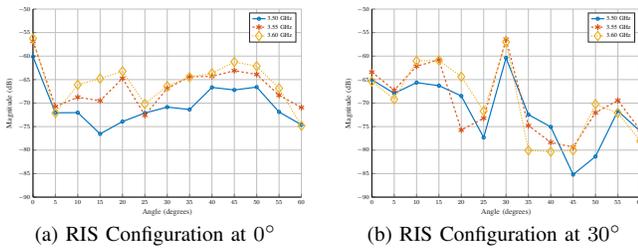

In this subsection, the proposed codebook is further examined in terms of frequency selectivity. Fig.~\ref{fig:freq_selectivity} analyzes the radiation patterns, extending the analysis to multiple frequencies. Specifically, we plot the patterns at 3.5 GHz, 3.55 GHz, and 3.6 GHz. While the angle direction of the main lobe is consistently maintained, highlighting the suitability of the codebook for beam sweeping and for wideband applications that require bandwidths of several MHz; we observe variations in the side lobes.

\subsection{Proposed Beam Sweeping Methodology for AoA estimation}
\begin{figure*}[t]
    \vspace{1mm}
    \centering
    \begin{minipage}{0.185\textwidth}
        \centering
        \scalebox{0.24}{
%
%
\definecolor{mycolor1}{rgb}{0.20000,0.60000,0.80000}%
\begin{tikzpicture}

\begin{axis}[%
width=4.844in,
height=3.396in,
at={(0.812in,0.458in)},
scale only axis,
bar shift auto,
xmin=-30,
xmax=0,
xlabel style={font=\color{white!15!black}},
xlabel={Normalized Power (dB)},
ymin=-0.2,
ymax=14.2,
ytick={ 1,  2,  3,  4,  5,  6,  7,  8,  9, 10, 11, 12, 13},
    yticklabels={$0^\circ$, $5^\circ$, $10^\circ$, $15^\circ$, $20^\circ$, $25^\circ$, $30^\circ$, $35^\circ$, $40^\circ$, $45^\circ$, $50^\circ$, $55^\circ$, $60^\circ$},
ylabel style={font=\color{white!15!black}},
ylabel={RIS Configuration},
axis background/.style={fill=white},
title style={font=\bfseries},
axis x line*=bottom,
axis y line*=left,
xmajorgrids,
ymajorgrids
]
\addplot[xbar, bar width=0.8, fill=mycolor1, draw=black, area legend] table[row sep=crcr] {%
0	1\\
-9.44701692725761	2\\
-20.4938110984389	3\\
-18.2215729458743	4\\
-19.3423000161992	5\\
-13.2961174665309	6\\
-6.65518493147796	7\\
-5.87754101561218	8\\
-6.98375988854762	9\\
-3.71912167808735	10\\
-10.3819347348705	11\\
-9.37497079929054	12\\
-19.3254123211246	13\\
};
\addplot[forget plot, color=white!15!black] table[row sep=crcr] {%
0	-0.2\\
0	14.2\\
};
\addplot[only marks, mark=*, mark options={}, mark size=2.2361pt, color=red, fill=red, forget plot] table[row sep=crcr]{%
x	y\\
0	1\\
};
\end{axis}
\end{tikzpicture}%
}
        \subcaption{Rx position at $0^\circ$}
        \label{fig:beamsw_out0}
    \end{minipage}
    \begin{minipage}{0.185\textwidth}
        \centering
        \scalebox{0.24}{
%
%
\definecolor{mycolor1}{rgb}{0.20000,0.60000,0.80000}%
\begin{tikzpicture}

\begin{axis}[%
width=4.844in,
height=3.396in,
at={(0.812in,0.458in)},
scale only axis,
bar shift auto,
xmin=-30,
xmax=0,
xlabel style={font=\color{white!15!black}},
xlabel={Normalized Power (dB)},
ymin=-0.2,
ymax=14.2,
ytick={ 1,  2,  3,  4,  5,  6,  7,  8,  9, 10, 11, 12, 13},
    yticklabels={$0^\circ$, $5^\circ$, $10^\circ$, $15^\circ$, $20^\circ$, $25^\circ$, $30^\circ$, $35^\circ$, $40^\circ$, $45^\circ$, $50^\circ$, $55^\circ$, $60^\circ$},
ylabel style={font=\color{white!15!black}},
ylabel={RIS Configuration},
axis background/.style={fill=white},
title style={font=\bfseries},
axis x line*=bottom,
axis y line*=left,
xmajorgrids,
ymajorgrids
]
\addplot[xbar, bar width=0.8, fill=mycolor1, draw=black, area legend] table[row sep=crcr] {%
-11.4319704177238	1\\
-13.9614533664424	2\\
-10.234631161402	3\\
0	4\\
-6.12230349350461	5\\
-4.83810975091131	6\\
-6.46597916579952	7\\
-12.241166064019	8\\
-13.0297536395889	9\\
-10.1663664046928	10\\
-14.0201408092067	11\\
-13.5358192784159	12\\
-13.4227979155706	13\\
};
\addplot[forget plot, color=white!15!black] table[row sep=crcr] {%
0	-0.2\\
0	14.2\\
};
\addplot[only marks, mark=*, mark options={}, mark size=2.2361pt, color=red, fill=red, forget plot] table[row sep=crcr]{%
x	y\\
0	4\\
};
\end{axis}
\end{tikzpicture}%
}
        \subcaption{Rx position at $15^\circ$}
        \label{fig:beamsw_out15}
    \end{minipage}
    \begin{minipage}{0.185\textwidth}
        \centering
        \scalebox{0.24}{
%
%
\definecolor{mycolor1}{rgb}{0.20000,0.60000,0.80000}%
\begin{tikzpicture}

\begin{axis}[%
width=4.844in,
height=3.396in,
at={(0.812in,0.458in)},
scale only axis,
bar shift auto,
xmin=-30,
xmax=0,
xlabel style={font=\color{white!15!black}},
xlabel={Normalized Power (dB)},
ymin=-0.2,
ymax=14.2,
ytick={ 1,  2,  3,  4,  5,  6,  7,  8,  9, 10, 11, 12, 13},
    yticklabels={$0^\circ$, $5^\circ$, $10^\circ$, $15^\circ$, $20^\circ$, $25^\circ$, $30^\circ$, $35^\circ$, $40^\circ$, $45^\circ$, $50^\circ$, $55^\circ$, $60^\circ$},
ylabel style={font=\color{white!15!black}},
ylabel={RIS Configuration},
axis background/.style={fill=white},
title style={font=\bfseries},
axis x line*=bottom,
axis y line*=left,
xmajorgrids,
ymajorgrids
]
\addplot[xbar, bar width=0.8, fill=mycolor1, draw=black, area legend] table[row sep=crcr] {%
-9.99352923435171	1\\
-7.60561512425468	2\\
-6.86197214320735	3\\
-8.51801882247999	4\\
-13.1258417597495	5\\
-18.0113336483307	6\\
0	7\\
-19.3672845909811	8\\
-17.2760209284037	9\\
-12.6625475217441	10\\
-20.1223729371069	11\\
-15.5199866971492	12\\
-10.2060252791208	13\\
};
\addplot[forget plot, color=white!15!black] table[row sep=crcr] {%
0	-0.2\\
0	14.2\\
};
\addplot[only marks, mark=*, mark options={}, mark size=2.2361pt, color=red, fill=red, forget plot] table[row sep=crcr]{%
x	y\\
0	7\\
};
\end{axis}
\end{tikzpicture}%
}
        \subcaption{Rx position at $30^\circ$}
        \label{fig:beamsw_out30}
    \end{minipage}
    \begin{minipage}{0.185\textwidth}
        \centering
        \scalebox{0.24}{
%
%
\definecolor{mycolor1}{rgb}{0.20000,0.60000,0.80000}%
\begin{tikzpicture}

\begin{axis}[%
width=4.844in,
height=3.396in,
at={(0.812in,0.458in)},
scale only axis,
bar shift auto,
xmin=-30,
xmax=0,
xlabel style={font=\color{white!15!black}},
xlabel={Normalized Power (dB)},
ymin=-0.2,
ymax=14.2,
ytick={ 1,  2,  3,  4,  5,  6,  7,  8,  9, 10, 11, 12, 13},
    yticklabels={$0^\circ$, $5^\circ$, $10^\circ$, $15^\circ$, $20^\circ$, $25^\circ$, $30^\circ$, $35^\circ$, $40^\circ$, $45^\circ$, $50^\circ$, $55^\circ$, $60^\circ$},
ylabel style={font=\color{white!15!black}},
ylabel={RIS Configuration},
axis background/.style={fill=white},
title style={font=\bfseries},
axis x line*=bottom,
axis y line*=left,
xmajorgrids,
ymajorgrids
]
\addplot[xbar, bar width=0.8, fill=mycolor1, draw=black, area legend] table[row sep=crcr] {%
-5.29345980933891	1\\
-7.62364946882121	2\\
-11.7327585397704	3\\
-15.1315223789185	4\\
-24.1553764203118	5\\
-12.2215085934111	6\\
-21.624419473844	7\\
-16.2499711696139	8\\
-4.52090925233889	9\\
0	10\\
-9.19676901356801	11\\
-12.5337483272285	12\\
-22.216815734731	13\\
};
\addplot[forget plot, color=white!15!black] table[row sep=crcr] {%
0	-0.2\\
0	14.2\\
};
\addplot[only marks, mark=*, mark options={}, mark size=2.2361pt, color=red, fill=red, forget plot] table[row sep=crcr]{%
x	y\\
0	10\\
};
\end{axis}
\end{tikzpicture}%
}
        \subcaption{Rx position at $45^\circ$}
        \label{fig:beamsw_out45}
    \end{minipage}
    \begin{minipage}{0.185\textwidth}
        \centering
        \scalebox{0.24}{
%
%
\definecolor{mycolor1}{rgb}{0.20000,0.60000,0.80000}%
\begin{tikzpicture}

\begin{axis}[%
width=4.844in,
height=3.396in,
at={(0.812in,0.458in)},
scale only axis,
bar shift auto,
xmin=-30,
xmax=0,
xlabel style={font=\color{white!15!black}},
xlabel={Normalized Power (dB)},
ymin=-0.2,
ymax=14.2,
ytick={ 1,  2,  3,  4,  5,  6,  7,  8,  9, 10, 11, 12, 13},
    yticklabels={$0^\circ$, $5^\circ$, $10^\circ$, $15^\circ$, $20^\circ$, $25^\circ$, $30^\circ$, $35^\circ$, $40^\circ$, $45^\circ$, $50^\circ$, $55^\circ$, $60^\circ$},
ylabel style={font=\color{white!15!black}},
ylabel={RIS Configuration},
axis background/.style={fill=white},
title style={font=\bfseries},
axis x line*=bottom,
axis y line*=left,
xmajorgrids,
ymajorgrids
]
\addplot[xbar, bar width=0.8, fill=mycolor1, draw=black, area legend] table[row sep=crcr] {%
-9.41516888312268	1\\
-14.1738385093149	2\\
-13.4048739907347	3\\
-16.1853162549615	4\\
-17.6233751281677	5\\
-13.7022503349284	6\\
-11.8131893581137	7\\
-16.3225709356877	8\\
-12.1850497581985	9\\
-15.4550154342765	10\\
-14.6098927014232	11\\
-3.36654273309183	12\\
0	13\\
};
\addplot[forget plot, color=white!15!black] table[row sep=crcr] {%
0	-0.2\\
0	14.2\\
};
\addplot[only marks, mark=*, mark options={}, mark size=2.2361pt, color=red, fill=red, forget plot] table[row sep=crcr]{%
x	y\\
0	13\\
};
\end{axis}
\end{tikzpicture}%
}
        \subcaption{Rx position at $60^\circ$}
        \label{fig:beamsw_out60}
    \end{minipage}
    \caption{Outdoor setup: Beam sweeping results}
    \label{fig:out_beamsw}
\end{figure*}
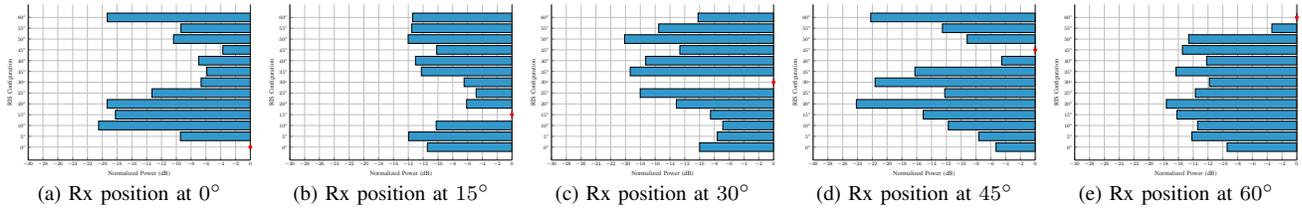

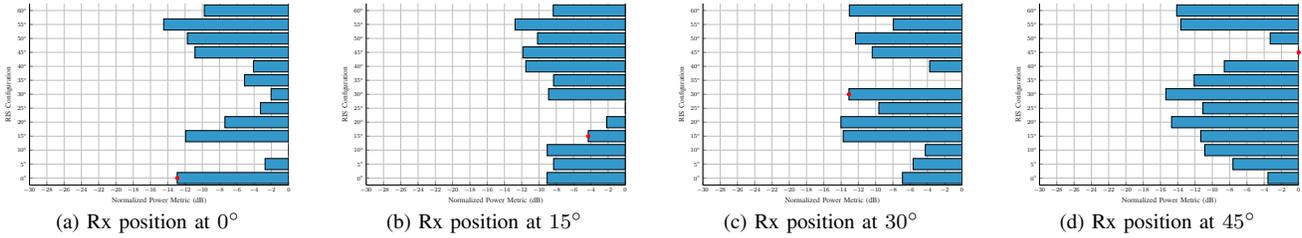
\begin{figure*}[t]
    \centering
    \begin{minipage}{0.24\textwidth}
        \centering
        \scalebox{0.28}{
%
%
\definecolor{mycolor1}{rgb}{0.20000,0.60000,0.80000}%
\begin{tikzpicture}

\begin{axis}[%
width=4.844in,
height=3.396in,
at={(0.812in,0.458in)},
scale only axis,
bar shift auto,
xmin=-30,
xmax=0,
xlabel style={font=\color{white!15!black}},
xlabel={Normalized Power Metric (dB)},
ymin=0.5,
ymax=13.5,
ytick={ 1,  2,  3,  4,  5,  6,  7,  8,  9, 10, 11, 12, 13},
    yticklabels={$0^\circ$, $5^\circ$, $10^\circ$, $15^\circ$, $20^\circ$, $25^\circ$, $30^\circ$, $35^\circ$, $40^\circ$, $45^\circ$, $50^\circ$, $55^\circ$, $60^\circ$},
ylabel style={font=\color{white!15!black}},
ylabel={RIS Configuration},
axis background/.style={fill=white},
title style={font=\bfseries},
axis x line*=bottom,
axis y line*=left,
xmajorgrids,
ymajorgrids
]
\addplot[xbar, bar width=0.8, fill=mycolor1, draw=black, area legend] table[row sep=crcr] {%
-12.9107723137461	1\\
-2.73862231211901	2\\
0	3\\
-11.9397246051296	4\\
-7.38691555900036	5\\
-3.26783917777293	6\\
-2.04516825088477	7\\
-5.11344900205867	8\\
-4.08550018629995	9\\
-10.8553865617718	10\\
-11.7329574611128	11\\
-14.4791219904472	12\\
-9.77057203755779	13\\
};
\addplot[forget plot, color=white!15!black, line width=1.5pt] table[row sep=crcr] {%
0	0.5\\
0	13.5\\
};
\addplot[only marks, mark=*, mark options={}, mark size=2.2361pt, color=red, fill=red, forget plot] table[row sep=crcr]{%
x	y\\
-12.9107723137461	1\\
};
\end{axis}
\end{tikzpicture}%
}
        \subcaption{Rx position at $0^\circ$}
        \label{fig:ind_beamsw_out0}
    \end{minipage}
    \begin{minipage}{0.24\textwidth}
        \centering
        \scalebox{0.28}{
%
%
\definecolor{mycolor1}{rgb}{0.20000,0.60000,0.80000}%
\begin{tikzpicture}

\begin{axis}[%
width=4.844in,
height=3.396in,
at={(0.812in,0.458in)},
scale only axis,
bar shift auto,
xmin=-30,
xmax=0,
xlabel style={font=\color{white!15!black}},
xlabel={Normalized Power Metric (dB)},
ymin=0.5,
ymax=13.5,
ytick={ 1,  2,  3,  4,  5,  6,  7,  8,  9, 10, 11, 12, 13},
    yticklabels={$0^\circ$, $5^\circ$, $10^\circ$, $15^\circ$, $20^\circ$, $25^\circ$, $30^\circ$, $35^\circ$, $40^\circ$, $45^\circ$, $50^\circ$, $55^\circ$, $60^\circ$},
ylabel style={font=\color{white!15!black}},
ylabel={RIS Configuration},
axis background/.style={fill=white},
title style={font=\bfseries},
axis x line*=bottom,
axis y line*=left,
xmajorgrids,
ymajorgrids
]
\addplot[xbar, bar width=0.8, fill=mycolor1, draw=black, area legend] table[row sep=crcr] {%
-9.05769334868475	1\\
-8.31192009670491	2\\
-9.04457383685525	3\\
-4.30040179904159	4\\
-2.15822465676875	5\\
0	6\\
-8.89203862886973	7\\
-8.31700676862815	8\\
-11.5286560284076	9\\
-11.8658155926617	10\\
-10.1694528255634	11\\
-12.7714223747898	12\\
-8.3503237107953	13\\
};
\addplot[forget plot, color=white!15!black, line width=1.5pt] table[row sep=crcr] {%
0	0.5\\
0	13.5\\
};
\addplot[only marks, mark=*, mark options={}, mark size=2.2361pt, color=red, fill=red, forget plot] table[row sep=crcr]{%
x	y\\
-4.30040179904159	4\\
};
\end{axis}
\end{tikzpicture}%
}
        \subcaption{Rx position at $15^\circ$}
        \label{fig:ind_beamsw_out15}
    \end{minipage}
    \begin{minipage}{0.24\textwidth}
        \centering
        \scalebox{0.28}{
%
%
\definecolor{mycolor1}{rgb}{0.20000,0.60000,0.80000}%
\begin{tikzpicture}

\begin{axis}[%
width=4.844in,
height=3.396in,
at={(0.812in,0.458in)},
scale only axis,
bar shift auto,
xmin=-30,
xmax=0,
xlabel style={font=\color{white!15!black}},
xlabel={Normalized Power Metric (dB)},
ymin=0.5,
ymax=13.5,
ytick={ 1,  2,  3,  4,  5,  6,  7,  8,  9, 10, 11, 12, 13},
    yticklabels={$0^\circ$, $5^\circ$, $10^\circ$, $15^\circ$, $20^\circ$, $25^\circ$, $30^\circ$, $35^\circ$, $40^\circ$, $45^\circ$, $50^\circ$, $55^\circ$, $60^\circ$},
ylabel style={font=\color{white!15!black}},
ylabel={RIS Configuration},
axis background/.style={fill=white},
title style={font=\bfseries},
axis x line*=bottom,
axis y line*=left,
xmajorgrids,
ymajorgrids
]
\addplot[xbar, bar width=0.8, fill=mycolor1, draw=black, area legend] table[row sep=crcr] {%
-6.8944050079383	1\\
-5.66031293307523	2\\
-4.25002119340891	3\\
-13.7551552384071	4\\
-14.0193744017144	5\\
-9.6291075319989	6\\
-13.0940070434133	7\\
0	8\\
-3.74509164205062	9\\
-10.3979020769042	10\\
-12.3303138725052	11\\
-7.94572172162777	12\\
-13.0489854492803	13\\
};
\addplot[forget plot, color=white!15!black, line width=1.5pt] table[row sep=crcr] {%
0	0.5\\
0	13.5\\
};
\addplot[only marks, mark=*, mark options={}, mark size=2.2361pt, color=red, fill=red, forget plot] table[row sep=crcr]{%
x	y\\
-13.0940070434133	7\\
};
\end{axis}
\end{tikzpicture}%
}
        \subcaption{Rx position at $30^\circ$}
        \label{fig:ind_beamsw_out30}
    \end{minipage}
    \begin{minipage}{0.24\textwidth}
        \centering
        \scalebox{0.28}{
%
%
\definecolor{mycolor1}{rgb}{0.20000,0.60000,0.80000}%
\begin{tikzpicture}

\begin{axis}[%
width=4.844in,
height=3.396in,
at={(0.812in,0.458in)},
scale only axis,
bar shift auto,
xmin=-30,
xmax=0,
xlabel style={font=\color{white!15!black}},
xlabel={Normalized Power Metric (dB)},
ymin=0.5,
ymax=13.5,
ytick={ 1,  2,  3,  4,  5,  6,  7,  8,  9, 10, 11, 12, 13},
    yticklabels={$0^\circ$, $5^\circ$, $10^\circ$, $15^\circ$, $20^\circ$, $25^\circ$, $30^\circ$, $35^\circ$, $40^\circ$, $45^\circ$, $50^\circ$, $55^\circ$, $60^\circ$},
ylabel style={font=\color{white!15!black}},
ylabel={RIS Configuration},
axis background/.style={fill=white},
title style={font=\bfseries},
axis x line*=bottom,
axis y line*=left,
xmajorgrids,
ymajorgrids
]
\addplot[xbar, bar width=0.8, fill=mycolor1, draw=black, area legend] table[row sep=crcr] {%
-3.56139278239337	1\\
-7.61564335874992	2\\
-10.8691034039982	3\\
-11.3367028591996	4\\
-14.7172716668008	5\\
-11.0924497633497	6\\
-15.3765862114877	7\\
-12.1016533771549	8\\
-8.61678803525348	9\\
0	10\\
-3.30239975627171	11\\
-13.6511576168329	12\\
-14.1264785918066	13\\
};
\addplot[forget plot, color=white!15!black, line width=1.5pt] table[row sep=crcr] {%
0	0.5\\
0	13.5\\
};
\addplot[only marks, mark=*, mark options={}, mark size=2.2361pt, color=red, fill=red, forget plot] table[row sep=crcr]{%
x	y\\
0	10\\
};
\end{axis}
\end{tikzpicture}%
}
        \subcaption{Rx position at $45^\circ$}
        \label{fig:ind_beamsw_out45}
    \end{minipage}
    
    \caption{Indoor setup: Beam sweeping results utilizing the precomputed codebook, with maximum AoA estimation error equal to $10^\circ$ (cases~\ref{fig:ind_beamsw_out0} and \ref{fig:ind_beamsw_out15}).}
    \label{fig:indoor_with_out}
\end{figure*}

\begin{figure*}[t]
    \centering
    \begin{minipage}{0.24\textwidth}
        \centering
        \scalebox{0.28}{
%
%
\definecolor{mycolor1}{rgb}{0.20000,0.60000,0.80000}%
\begin{tikzpicture}

\begin{axis}[%
width=4.844in,
height=3.396in,
at={(0.812in,0.458in)},
scale only axis,
bar shift auto,
xmin=-30,
xmax=0,
xlabel style={font=\color{white!15!black}},
xlabel={Normalized Power Metric (dB)},
ymin=0.5,
ymax=8.5,
ytick={1, 2, 3, 4, 5, 6, 7, 8},
    yticklabels={$0^\circ$-1st iter, $0^\circ$-2nd iter, $15^\circ$-1st iter, $15^\circ$-2nd iter, $30^\circ$-1st iter, $30^\circ$-2nd iter, $45^\circ$-1st iter, $45^\circ$-2nd iter},
ylabel style={font=\color{white!15!black}},
ylabel={RIS Configuration},
axis background/.style={fill=white},
title style={font=\bfseries},
axis x line*=bottom,
axis y line*=left,
xmajorgrids,
ymajorgrids
]
\addplot[xbar, bar width=0.8, fill=mycolor1, draw=black, area legend] table[row sep=crcr] {%
-0.338885874853034	1\\
0	2\\
-14.21294203993	3\\
-13.7635333493692	4\\
-10.140545719259	5\\
-9.09809874944966	6\\
-3.06616271812548	7\\
-3.36497427259179	8\\
};
\addplot[forget plot, color=white!15!black, line width=1.5pt] table[row sep=crcr] {%
0	0.5\\
0	8.5\\
};
\addplot[only marks, mark=*, mark options={}, mark size=2.2361pt, color=red, fill=red, forget plot] table[row sep=crcr]{%
x	y\\
0	2\\
};
\end{axis}
\end{tikzpicture}%
}
        \subcaption{Rx position at $0^\circ$}
        \label{fig:ind_beamsw_maxin0}
    \end{minipage}
    \begin{minipage}{0.24\textwidth}
        \centering
        \scalebox{0.28}{
%
%
\definecolor{mycolor1}{rgb}{0.20000,0.60000,0.80000}%
\begin{tikzpicture}

\begin{axis}[%
width=4.844in,
height=3.396in,
at={(0.812in,0.458in)},
scale only axis,
bar shift auto,
xmin=-30,
xmax=0,
xlabel style={font=\color{white!15!black}},
xlabel={Normalized Power Metric (dB)},
ymin=0.5,
ymax=8.5,
ytick={1, 2, 3, 4, 5, 6, 7, 8},
yticklabels={$0^\circ$-1st iter, $0^\circ$-2nd iter, $15^\circ$-1st iter, $15^\circ$-2nd iter, $30^\circ$-1st iter, $30^\circ$-2nd iter, $45^\circ$-1st iter, $45^\circ$-2nd iter},
ylabel style={font=\color{white!15!black}},
ylabel={RIS Configuration},
axis background/.style={fill=white},
title style={font=\bfseries},
axis x line*=bottom,
axis y line*=left,
xmajorgrids,
ymajorgrids
]
\addplot[xbar, bar width=0.8, fill=mycolor1, draw=black, area legend] table[row sep=crcr] {%
-11.7830591389383	1\\
-11.4120709740691	2\\
-0.143340601772081	3\\
0	4\\
-5.71565626489543	5\\
-5.12892298599902	6\\
-8.97111529264295	7\\
-6.56237911245133	8\\
};
\addplot[forget plot, color=white!15!black, line width=1.5pt] table[row sep=crcr] {%
0	0.5\\
0	8.5\\
};
\addplot[only marks, mark=*, mark options={}, mark size=2.2361pt, color=red, fill=red, forget plot] table[row sep=crcr]{%
x	y\\
0	4\\
};
\end{axis}
\end{tikzpicture}%
}
        \subcaption{Rx position at $15^\circ$}
        \label{fig:ind_beamsw_maxin15}
    \end{minipage}
    \begin{minipage}{0.24\textwidth}
        \centering
        \scalebox{0.28}{
%
%
\definecolor{mycolor1}{rgb}{0.20000,0.60000,0.80000}%
\begin{tikzpicture}

\begin{axis}[%
width=4.844in,
height=3.396in,
at={(0.812in,0.458in)},
scale only axis,
bar shift auto,
xmin=-30,
xmax=0,
xlabel style={font=\color{white!15!black}},
xlabel={Normalized Power Metric (dB)},
ymin=0.5,
ymax=8.5,
ytick={1, 2, 3, 4, 5, 6, 7, 8},
yticklabels={$0^\circ$-1st iter, $0^\circ$-2nd iter, $15^\circ$-1st iter, $15^\circ$-2nd iter, $30^\circ$-1st iter, $30^\circ$-2nd iter, $45^\circ$-1st iter, $45^\circ$-2nd iter},
ylabel style={font=\color{white!15!black}},
ylabel={RIS Configuration},
axis background/.style={fill=white},
title style={font=\bfseries},
axis x line*=bottom,
axis y line*=left,
xmajorgrids,
ymajorgrids
]
\addplot[xbar, bar width=0.8, fill=mycolor1, draw=black, area legend] table[row sep=crcr] {%
-3.4886760879522	1\\
-2.65544881564408	2\\
-3.30869249437968	3\\
-2.37971672279907	4\\
-0.665999262367848	5\\
0	6\\
-9.0090572620844	7\\
-6.70795064267047	8\\
};
\addplot[forget plot, color=white!15!black, line width=1.5pt] table[row sep=crcr] {%
0	0.5\\
0	8.5\\
};
\addplot[only marks, mark=*, mark options={}, mark size=2.2361pt, color=red, fill=red, forget plot] table[row sep=crcr]{%
x	y\\
0	6\\
};
\end{axis}
\end{tikzpicture}%
}
        \subcaption{Rx position at $30^\circ$}
        \label{fig:ind_beamsw_maxin30}
    \end{minipage}
    \begin{minipage}{0.24\textwidth}
        \centering
        \scalebox{0.28}{
%
%
\definecolor{mycolor1}{rgb}{0.20000,0.60000,0.80000}%
\begin{tikzpicture}

\begin{axis}[%
width=4.844in,
height=3.396in,
at={(0.812in,0.458in)},
scale only axis,
bar shift auto,
xmin=-30,
xmax=0,
xlabel style={font=\color{white!15!black}},
xlabel={Normalized Power Metric (dB)},
ymin=0.5,
ymax=8.5,
ytick={1, 2, 3, 4, 5, 6, 7, 8},
yticklabels={$0^\circ$-1st iter, $0^\circ$-2nd iter, $15^\circ$-1st iter, $15^\circ$-2nd iter, $30^\circ$-1st iter, $30^\circ$-2nd iter, $45^\circ$-1st iter, $45^\circ$-2nd iter},
ylabel style={font=\color{white!15!black}},
ylabel={RIS Configuration},
axis background/.style={fill=white},
title style={font=\bfseries},
axis x line*=bottom,
axis y line*=left,
xmajorgrids,
ymajorgrids
]
\addplot[xbar, bar width=0.8, fill=mycolor1, draw=black, area legend] table[row sep=crcr] {%
-4.19209448105993	1\\
-5.09574629996613	2\\
-12.113619763628	3\\
-15.6057824141141	4\\
-14.7521958130981	5\\
-11.4781886358824	6\\
-0.276869637951478	7\\
0	8\\
};
\addplot[forget plot, color=white!15!black, line width=1.5pt] table[row sep=crcr] {%
0	0.5\\
0	8.5\\
};
\addplot[only marks, mark=*, mark options={}, mark size=2.2361pt, color=red, fill=red, forget plot] table[row sep=crcr]{%
x	y\\
0	8\\
};
\end{axis}
\end{tikzpicture}%
}
        \subcaption{Rx position at $45^\circ$}
        \label{fig:ind_beamsw_maxin45}
    \end{minipage}
    
    \caption{Indoor setup: Beam sweeping results, executing two iterations of Algorithm~1 in real time (Maximization).}
    \label{fig:m1_indoor}
\end{figure*}
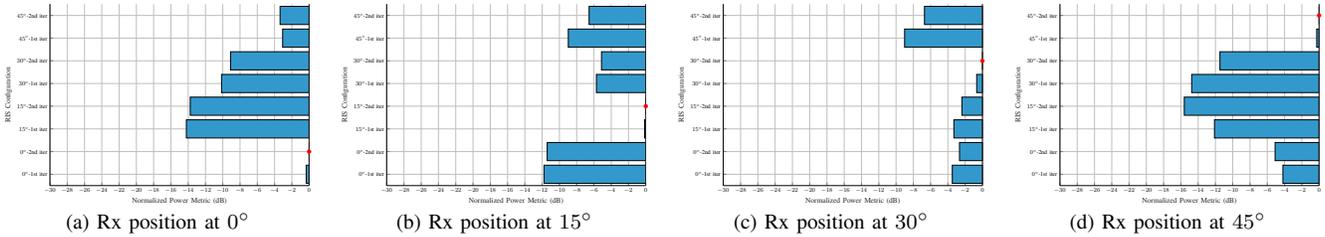

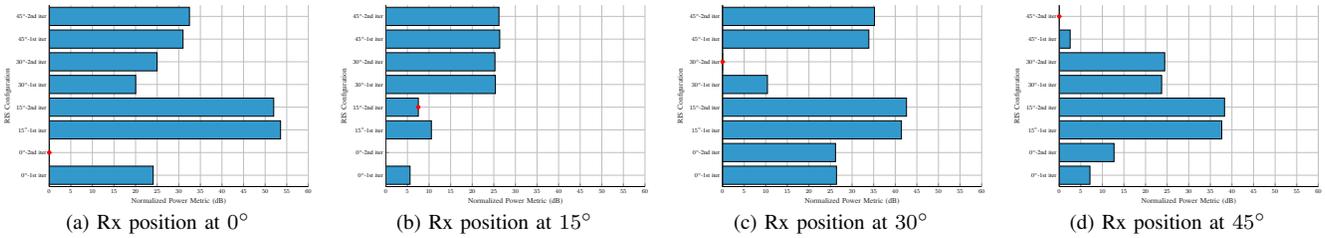
\begin{figure*}[t]
    \centering
    \begin{minipage}{0.24\textwidth}
        \centering
        \scalebox{0.28}{
%
%
\definecolor{mycolor1}{rgb}{0.20000,0.60000,0.80000}%
\begin{tikzpicture}

\begin{axis}[%
width=4.844in,
height=3.396in,
at={(0.812in,0.458in)},
scale only axis,
bar shift auto,
xmin=0,
xmax=60,
xlabel style={font=\color{white!15!black}},
xlabel={Normalized Power Metric (dB)},
ymin=0.5,
ymax=8.5,
ytick={1,2,3,4,5,6,7,8},
yticklabels={$0^\circ$-1st iter, $0^\circ$-2nd iter, $15^\circ$-1st iter, $15^\circ$-2nd iter, $30^\circ$-1st iter, $30^\circ$-2nd iter, $45^\circ$-1st iter, $45^\circ$-2nd iter},
ylabel style={font=\color{white!15!black}},
ylabel={RIS Configuration},
axis background/.style={fill=white},
title style={font=\bfseries},
axis x line*=bottom,
axis y line*=left,
xmajorgrids,
ymajorgrids
]
\addplot[xbar, bar width=0.8, fill=mycolor1, draw=black, area legend] table[row sep=crcr] {%
24.0764	1\\
0	2\\
53.56713	3\\
51.98386	4\\
20.0303	5\\
24.97505	6\\
30.9861	7\\
32.47019	8\\
};
\addplot[forget plot, color=white!15!black, line width=1.5pt] table[row sep=crcr] {%
0	0.5\\
0	8.5\\
};
\addplot[only marks, mark=*, mark options={}, mark size=2.2361pt, color=red, fill=red, forget plot] table[row sep=crcr]{%
x	y\\
0	2\\
};
\end{axis}
\end{tikzpicture}%
}
        \subcaption{Rx position at $0^\circ$}
        \label{fig:ind_beamsw_minin0}
    \end{minipage}
    \begin{minipage}{0.24\textwidth}
        \centering
        \scalebox{0.28}{
%
%
\definecolor{mycolor1}{rgb}{0.20000,0.60000,0.80000}%
\begin{tikzpicture}

\begin{axis}[%
width=4.844in,
height=3.396in,
at={(0.812in,0.458in)},
scale only axis,
bar shift auto,
xmin=0,
xmax=60,
xlabel style={font=\color{white!15!black}},
xlabel={Normalized Power Metric (dB)},
ymin=0.5,
ymax=8.5,
ytick={1,2,3,4,5,6,7,8},
yticklabels={$0^\circ$-1st iter, $0^\circ$-2nd iter, $15^\circ$-1st iter, $15^\circ$-2nd iter, $30^\circ$-1st iter, $30^\circ$-2nd iter, $45^\circ$-1st iter, $45^\circ$-2nd iter},
ylabel style={font=\color{white!15!black}},
ylabel={RIS Configuration},
axis background/.style={fill=white},
title style={font=\bfseries},
axis x line*=bottom,
axis y line*=left,
xmajorgrids,
ymajorgrids
]
\addplot[xbar, bar width=0.8, fill=mycolor1, draw=black, area legend] table[row sep=crcr] {%
5.60259000000001	1\\
0	2\\
10.54198	3\\
7.51519	4\\
25.36362	5\\
25.26755	6\\
26.36424	7\\
26.18726	8\\
};
\addplot[forget plot, color=white!15!black, line width=1.5pt] table[row sep=crcr] {%
0	0.5\\
0	8.5\\
};
\addplot[only marks, mark=*, mark options={}, mark size=2.2361pt, color=red, fill=red, forget plot] table[row sep=crcr]{%
x	y\\
7.51519	4\\
};
\end{axis}
\end{tikzpicture}%
}
        \subcaption{Rx position at $15^\circ$}
        \label{fig:ind_beamsw_minin15}
    \end{minipage}
    \begin{minipage}{0.24\textwidth}
        \centering
        \scalebox{0.28}{
%
%
\definecolor{mycolor1}{rgb}{0.20000,0.60000,0.80000}%
\begin{tikzpicture}

\begin{axis}[%
width=4.844in,
height=3.396in,
at={(0.812in,0.458in)},
scale only axis,
bar shift auto,
xmin=0,
xmax=60,
xlabel style={font=\color{white!15!black}},
xlabel={Normalized Power Metric (dB)},
ymin=0.5,
ymax=8.5,
ytick={1,2,3,4,5,6,7,8},
yticklabels={$0^\circ$-1st iter, $0^\circ$-2nd iter, $15^\circ$-1st iter, $15^\circ$-2nd iter, $30^\circ$-1st iter, $30^\circ$-2nd iter, $45^\circ$-1st iter, $45^\circ$-2nd iter},
ylabel style={font=\color{white!15!black}},
ylabel={RIS Configuration},
axis background/.style={fill=white},
title style={font=\bfseries},
axis x line*=bottom,
axis y line*=left,
xmajorgrids,
ymajorgrids
]
\addplot[xbar, bar width=0.8, fill=mycolor1, draw=black, area legend] table[row sep=crcr] {%
26.38895	1\\
26.16275	2\\
41.42751	3\\
42.59774	4\\
10.4021	5\\
0	6\\
33.87541	7\\
35.18893	8\\
};
\addplot[forget plot, color=white!15!black, line width=1.5pt] table[row sep=crcr] {%
0	0.5\\
0	8.5\\
};
\addplot[only marks, mark=*, mark options={}, mark size=2.2361pt, color=red, fill=red, forget plot] table[row sep=crcr]{%
x	y\\
0	6\\
};
\end{axis}
\end{tikzpicture}%
}
        \subcaption{Rx position at $30^\circ$}
        \label{fig:ind_beamsw_minin30}
    \end{minipage}
    \begin{minipage}{0.24\textwidth}
        \centering
        \scalebox{0.28}{
%
%
\definecolor{mycolor1}{rgb}{0.20000,0.60000,0.80000}%
\begin{tikzpicture}

\begin{axis}[%
width=4.844in,
height=3.396in,
at={(0.812in,0.458in)},
scale only axis,
bar shift auto,
xmin=0,
xmax=60,
xlabel style={font=\color{white!15!black}},
xlabel={Normalized Power Metric (dB)},
ymin=0.5,
ymax=8.5,
ytick={1,2,3,4,5,6,7,8},
yticklabels={$0^\circ$-1st iter, $0^\circ$-2nd iter, $15^\circ$-1st iter, $15^\circ$-2nd iter, $30^\circ$-1st iter, $30^\circ$-2nd iter, $45^\circ$-1st iter, $45^\circ$-2nd iter},
ylabel style={font=\color{white!15!black}},
ylabel={RIS Configuration},
axis background/.style={fill=white},
title style={font=\bfseries},
axis x line*=bottom,
axis y line*=left,
xmajorgrids,
ymajorgrids
]
\addplot[xbar, bar width=0.8, fill=mycolor1, draw=black, area legend] table[row sep=crcr] {%
7.12939999999999	1\\
12.7032	2\\
37.64182	3\\
38.28399	4\\
23.72995	5\\
24.439	6\\
2.54640000000001	7\\
0	8\\
};
\addplot[forget plot, color=white!15!black, line width=1.5pt] table[row sep=crcr] {%
0	0.5\\
0	8.5\\
};
\addplot[only marks, mark=*, mark options={}, mark size=2.2361pt, color=red, fill=red, forget plot] table[row sep=crcr]{%
x	y\\
0	8\\
};
\end{axis}
\end{tikzpicture}%
}
        \subcaption{Rx position at $45^\circ$}
        \label{fig:ind_beamsw_minin45}
    \end{minipage}
    
    \caption{Indoor setup: Beam sweeping results, executing two iterations of Algorithm~1 in real time (Minimization).}
    \label{fig:m3_indoor}
\end{figure*}

In this subsection, we evaluate beam sweeping as a method for AoA estimation in both outdoor and indoor experimental setups (Fig.~\ref{fig:exp_4images}). For these measurements, the Rx signal power is measured over a frequency range of 3.4 GHz to 3.6 GHz, sampled at 801 points using a VNA; further investigating the methodology's accuracy for wideband applications.
In Figs.~\ref{fig:out_beamsw}--\ref{fig:m3_indoor}
the red dot represents the RIS configuration where the maximum or minimum signal is expected. Furthermore, the Rx power is normalized, with the maximum set to 0 dB (Figs.~\ref{fig:out_beamsw}, \ref{fig:indoor_with_out}, and \ref{fig:m1_indoor}) or the minimum set to 0 dB (Fig.~\ref{fig:m3_indoor}).

\subsubsection{Beam Sweeping in Outdoor Setup}
As shown in Fig.~\ref{fig:out_beamsw}, the codebook generated by Algorithm~1 ensures a pronounced main lobe toward the desired Rx position, achieving at least a 5 dB gain over the second-best RIS configuration for all Rx positions. This highlights the robustness of the beamforming strategy; achieving perfect AoA estimation in the outdoor environment (limited multipath).

\subsubsection{Beam Sweeping in Indoor Setup}
Fig.~\ref{fig:indoor_with_out} illustrates the performance of the precomputed codebook, in the indoor setup, showing a maximum AoA estimation error of $10^\circ$. This aligns with the findings of~\cite{kompostiotis2024evaluation}, which reported similar errors in ray-tracing simulations,\linebreak and~\cite{rahal2023ris}, which corroborated these results in a real RIS FR2 setup. 
To refine the AoA estimation, we propose a methodology that employs RIS for both maximizing and minimizing the received signal power. This dual-mode operation highlights the diversity that the RIS can provide on measurements. For the maximization case, as shown in Fig.~\ref{fig:m1_indoor}, the real-time execution of Algorithm~1 achieves perfect AoA estimation accuracy. Regarding the minimization case, depicted in Fig.~\ref{fig:m3_indoor}, only one test exhibits a mismatch, while the others achieve perfect AoA estimation. This mismatch may be arised due to frequency selectivity, as variations in the side lobes are observed depending on the frequency (Fig.~\ref{fig:freq_selectivity}). As mentioned, the results are evaluated across the entire tested frequency range, and it is important to note that for specific frequencies, this mismatch does not occur; highlighting the impact of frequency-dependent effects on the beamforming performance. 
The real-time execution of Algorithm~1 allows for dynamic adjustments to the indoor environment's geometry, while both maximization and minimization beam sweeping modes provide complementary solutions, enhancing AoA estimation accuracy and offering a more robust approach for real-world indoor environments.

\vspace{-0.1cm}
\section{Conclusion}~\label{s:conclusion} 
This paper provides practical insights for the RIS technology, proposing a methodology for beam sweeping in both outdoor and indoor environments; addressing the challenges of multipath and frequency-selective behavior of RIS.
The applied RIS configuration method leads to a codebook that is practical for beam sweeping, and the real measurements validate the theoretical model. Outdoor measurements show that the generated codebook performs effectively, while for the challenging indoor setup, an additional methodology is proposed for improved AoA estimation, reducing the initial mismatch of around $10^\circ$ by adapting the RIS configuration to the specific multipath conditions and geometry of the environment.


\vspace{-0.1cm}
\section{Acknowledgement}
This work has been supported by the ESA Project PRISM: RIS-enabled Positioning and Mapping (NAVISP-EL1-063).

\bibliographystyle{IEEEtran}
\bibliography{irs}

\end{document}